\documentclass[11pt]{article}
\usepackage{amsmath}
\usepackage{amssymb}
\usepackage{theorem}
\newtheorem{proposition}{Proposition}[section]

\newtheorem{theorem}{Theorem}[section]
\newtheorem{conjecture}{Conjecture}[section]
\newtheorem{lemma}{Lemma}[section]
\newtheorem{remark}{Remark}[section]
\newtheorem{example}{Example}[section]
\newtheorem{corollary}{Corollary}[section]
\newcommand{\N}{\mathbb{N}}
\newcommand{\Z}{\mathbb{Z}}
\renewcommand{\P}{\underline{P}}
\renewcommand{\S}{\underline{S}}
\newcommand{\C}{\underline{C}}
\newcommand{\A}{\underline{A}}
\def\petitcarre{\vrule height4pt width 4pt depth0pt}
\def\enddim{\relax\ifmmode\eqno{\hbox{\petitcarre}}
\else
{\unskip\nobreak\hfil\penalty50
   \hskip2em\hbox{}\nobreak\hfil
   \petitcarre
   \parfillskip=0pt \finalhyphendemerits=0
  \par\medskip}\fi}

\def \begdim {\noindent {\sc Proof} : \par \noindent}

\DeclareMathOperator{\supp}{supp}
\begin{document}

\date{}
%\numberwithin{theorem}{subsection}

\numberwithin{equation}{section}
\title{\Large \bf
A note on
the factorization
conjecture
\thanks{
Partially supported by
the $FARB$ Project
{\it ``Aspetti computazionali e propriet\`a
algebriche degli automi e dei linguaggi formali''}
(University of Salerno, 2011), the
$FARB$ Project
{\it ``Aspetti algebrici e computazionali nella teoria dei codici
e dei linguaggi formali''}
(University of Salerno, 2012)
and the $MIUR$ Project $2010$-$2011$
{\it ``Automata and Formal Languages: Mathematical and Applicative Aspects''.}}}
\author{Clelia De Felice \\
Dipartimento di Informatica, \\
Universit\`a degli Studi di Salerno, \\
via Giovanni Paolo II, 132\\
84084 Fisciano (SA), Italy \\
defelice@dia.unisa.it}

\maketitle

%------------
\begin{abstract}
We give partial results on
the factorization conjecture on codes
proposed by Sch\"{u}tzenberger.
We consider a family of
finite maximal codes $C$ over the alphabet
$A = \{a, b\}$ and we prove that the factorization
conjecture holds for these codes.
This family contains $(p,4)$-codes, where a $(p,4)$-code
$C$ is a finite maximal code over $A$ such that each word in $C$
has at most four occurrences of $b$
and $a^p \in C$, for a prime number $p$.
We also discuss the structure of these
codes. The obtained results once again show relations
between factorizations of finite maximal codes
and factorizations of finite cyclic groups.
\end{abstract}
%----------------------

\section{Introduction} \label{introduction}

The theory of {\it variable-length codes} is a topic
with elegant mathematical results and strong
connections with automata theory. The theory originated
at the end of the $60$'s with Sch\"{u}tzenberger,
who proposed in \cite{SC55} the semigroup theory as a
mathematical setting for the
study of the uniquely decipherable
sets of words in the context
of information theory (see \cite{BPR} for
a complete treatment of this topic and also
\cite{IT} for a viewpoint focused on applications
of codes). In this paper
we follow this algebraic approach and codes
are defined as the {\it bases} of the free
submonoids of a free monoid.

A well known class of codes is that
of {\it prefix} codes, i.e., codes such that
none of their words is a left factor of another.
A classical representation of a finite prefix code $C$ over
an alphabet $A$ is as a set
of leaves on a tree. In this case, the set of the internal
nodes represents the set of the proper left factors
$P$ of $C$ and $C$ is {\it maximal} (i.e., $C$ is not properly
contained in any other code over $A$)
if and only if each internal node has a number of children
equal to the cardinality of $A$. Thus, it is clearly evident
that $C = PA \setminus P$ or, in terms of noncommutative polynomials,
$\underline{C} - 1 = \underline{P}(\underline{A} - 1)$
(here $1$ is the empty word and \underline{X} denotes
the {\it characteristic polynomial}
of a finite language $X$, i.e., the formal sum of its elements).

One of the conjectures proposed by Sch\"{u}tzenberger,
known as the {\it factorization conjecture}, asks whether
a more general equation can be stated for a finite
(not necessarily prefix) maximal code $C$, namely
whether finite subsets $P$, $S$ of $A^*$ exist such that
$\underline{C} - 1 = \underline{P}(\underline{A} - 1)\underline{S}$
\cite{P1,SC}.
This longstanding open question,
one of the most important in the theory of
codes, is inspired by a problem of information theory \cite{PS77}.

Only partial results are known (see \cite{BPR}).
The major contribution to this conjecture is due to
Reutenauer \cite{REU83,REU85}. In particular,
he proved that for any finite
maximal code $C$ over $A$, there exist polynomials
$P, S \in \Z \langle A \rangle$ such that
$\underline{C} - 1 = P(\underline{A} - 1)S$.
We call $(P,S)$ a {\it factorization} for $C$. Moreover
we say that a factorization $(P,S)$ for $C$ is
{\it positive} if
$P,S$ or $-P, -S$ have coefficients $0,1$.\footnote[1]{Note that in this paper we use
the term ``positive factorization'' with a slightly different
meaning with respect to the definition of the same term
in \cite{BPR}.}

Positive factorizations always exist
for a finite maximal code $C$ over a one-letter alphabet
and these positive factorizations have all been constructed in
\cite{KRR}. However, factorizations
which are not positive also exist, even for these simple codes.
In the case of a one-letter alphabet,
it has also been conjectured that if $(P,S)$ is a
factorization for $C$ and $S$ has coefficients $0,1$,
then the same holds for $P$ \cite{KH}.

There are not many examples of factorizations which are
not positive. On the contrary, {\it every} factorization
for $C$ is positive
if $C$ is a finite maximal code
over a two-letter alphabet $\{a,b\}$ with
$m \leq 3$ occurrences of the letter $b$ in
its words \cite{DF93,DFR,RES}.
In this paper, we investigate this further
and we prove the results which are briefly explained
below.

Let $A = \{a, b\}$.
For a polynomial $S \in \Z \langle  A \rangle$,  we denote
by $\supp(S)$ the set of words in  $A^*$
having a non-zero coefficient in $S$.
Let $C$ be a finite maximal code over $A$ such that $a^p \in C$,
for a prime number $p$.
Let $(P, S)$ be a factorization for $C$ such that,
for any word $w \in A^*$,
if $wba^j$ is in $\supp(S)$
then $a^j$ is also in $\supp(S)$.

First, we show that if $S \in \N \langle A \rangle$, then $(P,S)$ is positive
(Theorem \ref{mainteoremGen}).
This result is related to the above-mentioned
conjecture in \cite{KH}.
Second, we prove that if $\supp(S) \subseteq a^* \cup a^*ba^*$ then $(P,S)$
is positive (Theorem \ref{mainteorem1}).
Moreover, in this case we may inductively construct
all these factorizations $(P,S)$ (Section \ref{constructionC}).

A  $(p,4)$-code $C$ is a finite maximal code over $A$
containing $a^p$ and such that each word in $C$
has at most four occurrences of $b$.
A corollary of the previous results is that if $C$
is a $(p,4)$-code, for a prime number $p$, then each factorization
for $C$ is positive (Theorem \ref{mainteorem2}).

Finally, for
a polynomial $P \in \Z \langle  A \rangle$, let
$P_g$ be polynomials such that a word $w \in A^*$
has a non-zero coefficient $\alpha$ in $P_g$ if
and only if $w$
has $g$ occurrences of the letter $b$ and $w$
has the same non-zero coefficient $\alpha$ in $P$.
Let $(P,S)$ be a factorization
for a finite maximal code $C$ over $\{a, b \}$
such that if $a^iba^j$ is in $\supp(S_1)$ then
$a^j$ is in $\supp(S_0)$.
We prove that if $P_0, S_0, S_1$ have
coefficients $0, 1$, then $P_1$ has nonnegative coefficients
(Theorem \ref{mainteorem3}).

Another objective is the description
of the structure of the {\it (positively) factorizing}
codes, i.e., codes satisfying the factorization
conjecture. There are several papers
devoted to this problem \cite{DF89,DF93,DF96,DF01,DF05,DF06,DF07a,DF07b,RES}.
In particular, the structure of $m$-codes, $m \leq 3$,
has been characterized, as well as that
of codes $C$ such that $\C = \P(\A-1)\S + 1$, with
$P \subseteq A^*$, $S \subseteq a^*$.
In all these cases, there are relations
between (positive) factorizations of finite
maximal codes and factorizations of cyclic groups.
We tackle this problem for
$(p,4)$-codes and
the results proved in this paper once again show these relations.

The paper is organized as follows.
In Section \ref{Ba}, we set up
the basic definitions
and known results we need.
In Section \ref{outline}, we give an outline of the results
on the factorization conjecture and we prove
these results in Section \ref{main}.
In Section \ref{outline1}, we recall some known results
and give an outline of new results
on positively factorizing codes. The new results will be
stated in Sections \ref{construction1},  \ref{constructionC} and
\ref{construction4}.
Finally, in
Section \ref{finale}, we discuss some
open problems that follow on from these results.

%---------------------------------------

\section{Basics} \label{Ba}

\subsection{Codes and words}

Let $A^{*}$ be the {\it free monoid}
generated by a finite alphabet $A$
and let $A^+=A^{*} \setminus 1$ where $1$ is
the empty word.
For a word $w \in A^*$ and a
letter $a \in A$, we denote by $|w|$ the {\it length}
of $w$ and by $|w|_a$ the number of the occurrences
of $a$ in $w$.
The {\it reversal} of a word $w = a_1 \ldots a_n$,
$a_i \in A$, is the word $w^{\sim} = a_n \ldots a_1$
and we set $X^{\sim}=\{w^{\sim} ~|~ w \in X \}$.

A {\it code} $C$ is a subset
of $A^{*}$ such that, for all $h, k \geq 0$ and
$c_1, \ldots , c_h, c'_1, \ldots , c'_k \in C$, we have
\[
c_1 \cdots c_h= c'_1 \cdots c'_k \quad \Rightarrow \quad
h=k \quad \mbox{and} \quad c_i=c'_i
\quad \mbox{for} \quad i = 1, \ldots , h.
\]
A set $C \subseteq A^+$, such that $C \cap CA^+ = \emptyset$,
is a {\it prefix} code. $C$ is a {\it suffix} code if
$C^{\sim}$ is a prefix code and $C$ is a {\it biprefix}
code when
$C$ is both a suffix and a prefix code.
A code $C$ is a {\it maximal} code over $A$ if for each code $C'$
over $A$ such that $C \subseteq C'$ we have $C=C'$.

\subsection{Polynomials}

Let $\Z \langle A \rangle$
(resp. $\N \langle A \rangle$)
denote the semiring of the {\it
polynomials} with noncommutative variables in $A$
and integer (resp. nonnegative integer)
coefficients.
For a finite subset
$X$ of $A^{*}$,  $\underline{X}$ denotes its
{\it characteristic polynomial}, defined by
$\underline{X}= \sum_{x \in X} x$.
Therefore, ``characteristic polynomial'' will be synonymous
with ``polynomial with coefficients $0,1$''.
For a polynomial
$P$ and a word $w \in A^{*}$, $(P,w)$
denotes the coefficient of $w$ in $P$ and
we set $\supp(P) = \{ w \in A^* ~|~ (P,w) \not= 0\}$.
If $\supp(P) = \emptyset$, then $P = 0$ is the null polynomial.
When we write $P \geq Q$, with $P,Q \in \Z \langle A \rangle$,
we mean that $(P,w) \geq (Q, w)$, for any $w \in A^*$.
In particular, $P \geq 0$ means that
$P \in \N \langle A \rangle$.
Furthermore, $P^{\sim}$ is defined by
$(P^{\sim}, w^{\sim}) = (P,w)$, for each $w \in A^*$.
For $P \in \Z \langle A \rangle$, $A=\{a,b\}$ and
$g \in \N$, we
denote by $P_g$ polynomials such that
$$\forall w \in A^* \quad
(P_g, w) = \begin{cases} (P, w) & \text{if $|w|_b = g$}, \\
0 & \text{otherwise.} \end{cases}$$
Then for any $P \in \Z \langle A \rangle$,
there exists $h \in \N$ such that
$P=P_0 + \ldots + P_h$.
We write, as usual, $\Z[a]$ and $\N[a]$ instead of
$\Z \langle a \rangle$  and $\N \langle a \rangle$.
The map which associates the polynomial $\sum_{n \in \N} (H,n)a^{n} \in \N[a]$
to a finite multiset $H$ of
nonnegative integers, is a bijection between
the set of the finite multisets $H$ of nonnegative integers
and $\N[a]$.
We represent this bijection by the notation
$a^H = \sum_{n \in \N} (H,n)a^{n}$.
For example, $a^{\{0, 0, 1, 1, 1, 3, 4 \}} = 2 + 3a + a^3 + a^4$.
Consequently, the following computation rules are defined:
$a^{M+L}=a^{M}a^{L}$,
$a^{M \cup L}= a^{M} + a^{L}$,
$a^{\emptyset}=0$, $a^0=1$.

\subsection{Factorization conjecture}

Conjecture \ref{FC}, given in a weaker form
in \cite{PS77},
is among the most difficult, unsolved
problems in the theory of codes. This
conjecture was formulated by
Sch\"{u}tzenberger but, as
far as we know, it does not appear
explicitly in any of his papers.
It was quoted as the
{\it factorization conjecture} in \cite{P1}
for the first time and then also reported
in \cite{BPR,BER,BRLA}.
\begin{conjecture} \cite{SC} \label{FC}
Given a finite maximal code
$C$, there are finite subsets $P$, $S$ of $A^*$
such that:
\begin{eqnarray*}
\C -1= \P(\A -1)\S.
\end{eqnarray*}
\end{conjecture}
Each code $C$ verifying the
previous conjecture is finite, maximal
and is called a {\it (positively) factorizing code}.

Finite maximal prefix codes are the simplest
examples of positively factorizing codes.
Indeed, $C$ is a finite maximal prefix code if and
only if $\C = \P(\A-1) + 1$ for a finite subset
$P$ of $A^*$ \cite{BPR}.
In the previous relation,
$P$ is the set of the proper prefixes
of the words in $C$.
More interesting constructions
of factorizing codes can be found in
\cite{BO78,BO79,BO81},
whereas the result which is closest to
a solution of the conjecture
is reported in Theorem \ref{ine}
and was obtained by Reutenauer
\cite{BPR,BER,REU83,REU85}.

\begin{theorem} \label{ine} \cite{REU85}
Let $C \in \N \langle A \rangle$,
with $(C,1)=0$, and let
$P,S \in \Z  \langle A \rangle$ be
such that $C=P(\A-1)S+1$. Then, $C$ is the characteristic polynomial
of a finite maximal code. Furthermore, if
$P,S \in \N \langle A \rangle$,
then $P,S$ are polynomials with coefficients $0,1$.
Conversely, for any finite maximal code $C$
there exist
$P,S \in \Z \langle A \rangle$
such that $\C=P(\A-1)S+1$.
\end{theorem}

\noindent
Given a finite maximal code $C$, a {\it factorization}
$(P,S)$ for $C$ is a pair of polynomials
$P,S \in \Z \langle A \rangle$
such that $\C=P(\A-1)S+1$. Of course,
$(P,S)$ is a factorization for $C$ if and
only if the same holds for $(-P, -S)$
and, moreover, $(S^{\sim}, P^{\sim})$
is a factorization for $C^{\sim}$.
We say that a factorization $(P,S)$ for $C$ is
{\it positive} if
$P,S$ or $-P, -S$ have coefficients $0,1$.
From now on, $A=\{a,b\}$ will be a
two-letter alphabet.

\section{Outline of the results on the factorization
conjecture} \label{outline}

Let $C$ be a finite maximal code over $A$,
let $(P,S)$ be a factorization for $C$
(Theorem \ref{ine}).
Then $P,S \in \Z \langle A \rangle$
are such that $\C=P(\A-1)S+1$.
Thus, the characteristic polynomial $\C_r$
of the set $C_r = \{w \in C ~|~ |w|_b = r \}$
of the words in $C$ with $r$ occurrences of $b$, is
the sum of the terms of degree $r$ with respect
to the variable $b$ in the polynomial
$P(\A-1)S +1$, i.e.,
\begin{eqnarray}  \label{eqCodes0}
\C_{0} & = & P_{0}(a-1)S_{0}+1,
\end{eqnarray}
\begin{eqnarray} \label{eqCodesR}
\forall r \geq 0 \quad
\C_{r+1} & = &
\sum_{i+j=r}P_{i}bS_{j} + \sum_{i+j=r+1}P_{i}(a-1)S_{j}.
\end{eqnarray}
\begin{example} \label{EX1}
{\rm Consider the finite maximal code defined by the relation
$\C  = P(\A-1)S + 1$, with
\begin{eqnarray*}
P & = & 1 + a^{2}ba^{\{0,1,2,3,4,5,6\}} + a^{2}ba^3ba^{\{0,1,2,3,4,5,6\}}, \\
S & = & a^{\{0,1,2,3,4\}} + a^{\{ 0,1 \}}ba^{\{0,1,2,3,4\}}.
\end{eqnarray*}
Then we have}
\begin{eqnarray*}
P_0 &=& 1, \quad P_1 = a^{2}ba^{\{0,1,2,3,4,5,6\}}, \quad
P_2 = a^{2}ba^3ba^{\{0,1,2,3,4,5,6\}}, \\
S_0 &=& a^{\{0,1,2,3,4\}}, \quad S_1 = a^{\{ 0,1 \}}ba^{\{0,1,2,3,4\}}.
\end{eqnarray*}
{\rm Consequently}
\begin{eqnarray*}
\C_0 &=& P_0 (a-1) S_0 + 1 = (a-1)a^{\{0,1,2,3,4\}} + 1 = a^{5}, \\
\C_1 &=& P_0 b S_0 + P_1(a-1)S_0 + P_0(a-1)S_1 \\
    &=& ba^{\{0,1,2,3,4\}} +
a^{2}ba^{\{0,1,2,3,4, 5, 6\}}(a-1)a^{\{0,1,2,3,4\}} +
     (a-1)a^{\{ 0,1 \}}ba^{\{0,1,2,3,4\}} \\
     & = & a^{2}ba^{\{7,8,9,10,11\}}, \\
\C_2 &=& P_1 b S_0 + P_0bS_1 + P_1(a-1)S_1 + P_2(a-1)S_0 \\
    &=& a^{2}ba^{\{0,1,2,3,4,5,6\}}ba^{\{0,1,2,3,4\}}
    + b a^{\{ 0,1 \}}ba^{\{0,1,2,3,4\}} + \\
    & & a^{2}ba^{\{0,1,2,3,4,5,6\}}(a-1)a^{\{ 0,1 \}}ba^{\{0,1,2,3,4\}}
    + a^{2}ba^3ba^{\{0,1,2,3,4,5,6\}}(a-1)a^{\{0,1,2,3,4\}} \\
    &=& b a^{\{ 0,1 \}}ba^{\{0,1,2,3,4\}} +
    a^{2}ba^{\{2,4,5,6,7,8\}}ba^{\{0,1,2,3,4\}} +
    a^{2}ba^3ba^{\{7,8,9,10, 11\}}, \\
\C_3 &=& P_1bS_1 + P_2 b S_0 + P_2(a-1)S_1  \\
    &=& a^{2}ba^{\{0,1,2,3,4,5,6\}}ba^{\{ 0,1 \}}ba^{\{0,1,2,3,4\}} +
    a^{2}ba^3ba^{\{0,1,2,3,4,5,6\}}ba^{\{0,1,2,3,4\}} + \\
    & &  a^{2}ba^3ba^{\{0,1,2,3,4,5,6\}}(a-1)a^{\{ 0,1 \}}ba^{\{0,1,2,3,4\}} \\
    &=& a^{2}ba^{\{0,1,2,3,4,5,6\}}ba^{\{ 0,1 \}}ba^{\{0,1,2,3,4\}} +
    a^{2}ba^3ba^{\{2,3,4,5,6,7,8 \}}ba^{\{0,1,2,3,4\}}, \\
\C_4 &=& P_2bS_1 = a^{2}ba^3ba^{\{0,1,2,3,4,5,6\}} b a^{\{ 0,1 \}}b a^{\{0,1,2,3,4\}}.
\end{eqnarray*}
\end{example}

The factorization $(P,S)$ for $C$ in Example \ref{EX1} is positive.
We notice that $a^p \in C$, with $p$ a prime number,
$S = S_0 + S_1$, and words $a^iba^j$ in $\supp(S)$ are such that
$a^j$ is also in $\supp(S)$.
We will prove that {\it each} factorization $(P,S)$,
for a code $C$ satisfying these
hypotheses, is necessarily positive.

In the proof of this result
we may assume that $P = P_0 + \ldots + P_k$ with $k \geq 2$ and
$S = S_0 + S_1$ with $S_1 \not = 0$. Indeed, we recall below that
$(P,S)$ is always positive if $k \leq 1$ (and $S = S_0 + S_1$)
or if $S \in \Z[a]$.
Given $m \in \N$,
an $m$-code $C$ is a finite maximal
code over $\{a,b\}$ such that each word in $C$
has at most $m$ occurrences of $b$, and at least one word of $C$
contains exactly $m$ occurrences of $b$.
The following result has been proved in \cite{RES} for
$m = 1$, in \cite{DFR} for $m = 2$ and in \cite{DF93}
for $m = 3$.

\begin{theorem} \label{123}
Let $m \in \N$, $m \leq 3$.
Any $m$-code $C$ is
positively factorizing.
Moreover, each factorization for $C$ is
a positive factorization for $C$.
\end{theorem}

\noindent
The following results have been proved in \cite{DF93}.

\begin{theorem} \label{P(a)}
Let $C$ be a finite maximal code, let
$(P, S)$ be a factorization for $C$.
If $P \in \Z \langle a \rangle$ or
$S \in \Z  \langle a \rangle$,
then $(P, S)$ is a positive factorization
for $C$.
\end{theorem}

\begin{theorem} \label{T2}
Let $C$ be a finite maximal code, let
$(U,V)$ be a factorization for $C$.
Then either $(P,S) = (U,V)$ or $(P,S) = (-U,-V)$ satisfies
the following conditions, where  $P = \sum_{i=0}^{k} P_{i}$,
$S = \sum_{i=0}^{h} S_{i}$.
\begin{itemize}
\item[(i)]
$P_{k}$ and $S_{h}$ have coefficients $0,1$.
\item[(ii)]
$P_{k-1} \in \N \langle A \rangle
\setminus \{0 \}$ and there are finite sets $L_p$ of nonnegative integers,
for $p \in \supp(P_{k-1})$,  such that
$P_{k} = \sum_{p \in \supp(P_{k-1})} pba^{L_{p}}$ or
$S_{h-1} \in \N \langle A \rangle \setminus \{0 \}$
and there are finite sets $M_s$ of nonnegative integers,
for $s \in \supp(S_{h-1})$, such that
$S_{h} = \sum_{s \in \supp(S_{h-1})}a^{M_{s}}bs$.
\end{itemize}
\end{theorem}

Theorem \ref{T2} states that
$P_k$, $S_1$ or $-P_k$, $-S_1$ always have coefficients $0,1$.
If $a^p \in C$, we have
$\C_{0} = a^p = P_{0}(a-1)S_{0}+1$,
i.e., $P_0 S_0 = 1 + a + \ldots + a^{p-1}$.
Since for a prime number $p$,
the polynomial $1 + a + \ldots + a^{p-1}$
is irreducible in $\Z [a]$
(see Example, p. 129 in \cite{Lang}),
one of the pairs $(P_0, S_0)$, $(S_0, P_0)$, $(-P_0, -S_0)$,
$(-S_0, -P_0)$ is equal to the pair
$(1, 1 + a + \ldots + a^{p-1})$.
In Section \ref{main1}, we will prove that if
$S_1$ has
coefficients $0,1$, then the same holds for polynomials
$P_0, S_0$ (Lemma \ref{LL0}).

In conclusion, $P_0$ and $S = S_0 + S_1$ or $-P_0$ and $-S$ have coefficients $0,1$.
Next, we will consider factorizations $(P,S)$ for $C$, with $a^p \in C$
and where $S = S_0 + \ldots + S_t$, $t \geq 1$,
is such that, for any word $w$, if  $wba^j \in \supp(S)$
then $a^j \in \supp(S_0)$.
In Section \ref{main1}, we will prove that if $S \in \N \langle A \rangle$,
then $P$ is also in $\N \langle A \rangle$
(Theorem \ref{mainteoremGen}).
Hence, $P,S$ have coefficients $0,1$ (Theorem \ref{ine}).
As a consequence, we state our main result:
if $S = S_0 + S_1$, then $S \in \N \langle A \rangle$ and
$(P,S)$ is positive (Theorem \ref{mainteorem1}).

Regarding the factorization conjecture, we will
prove another result.
In Section \ref{main2}, we consider $(p,4)$-codes,
i.e., $4$-codes $C$ such that
$C \cap a^* = \{a^p\}$.
If $(P,S)$ is a
factorization for $C$
such that neither $P$ nor $S$ is in $\Z[a]$,
then either $P = P_0 + P_1 + P_2$, $S = S_0 + S_1$ with
$P_2 \not = 0$ and $S_1 \not = 0$ or
$P = P_0 + P_1$, $S = S_0 + S_1 + S_2$ with
$P_1 \not = 0$ and $S_2 \not = 0$ (see Example \ref{EX1}).
Assume $S = S_0 + S_1$.
One of the two cases in item (ii) of Theorem \ref{T2} applies to $(P,S)$.
In the second of these cases, the above-mentioned results show that $(P,S)$ is positive.
We will easily prove that $(P,S)$ is positive in the first case also.
The same arguments apply if $P = P_0 + P_1$.
Therefore, we show that all $(p,4)$-codes
have only positive factorizations.
Notice that in \cite{ZHG} it has been proved that
an $m$-code $C$
is positively factorizing if $b^m \in C$ and $m$ is a
prime number or $m=4$.

Finally, in Section \ref{main3} we will prove
that $P_1$ has nonnegative coefficients under weaker
hypotheses on $P,S,C$.
More precisely, we remove the hypothesis on the power of $a$ in $C$.
We assume that $(P,S)$ is a factorization for $C$ such that
if $a^iba^j$ is in $\supp(S_1)$ then
$a^j$ is in $\supp(S_0)$.
We prove that if $P_0, S_0, S_1$ have
coefficients $0, 1$, then $P_1$ has nonnegative coefficients
(Theorem \ref{mainteorem3}).
In the proof of this result, we point out properties
of $P_1$ and $S_1$ that will be used in
Sections \ref{construction1} and \ref{construction4}
for the construction of factorizing codes.

\section{Main results}  \label{main}

\subsection{Factorizations $(P,S)$ with one $b$ in $S$} \label{main1}

In this section, we prove our main result. We consider
a factorization $(P,S)$ for a
finite maximal code $C$ over
$A$, with $C \cap a^* = \{a^p\}$,
for a prime number $p$, and
$S = S_0 + S_1$. We assume that $a^iba^j \in \supp(S_1)$
implies $a^j \in \supp(S_0)$.
We prove that $(P,S)$ is positive
(Theorem \ref{mainteorem1}).
This result is a direct consequence of
Lemma \ref{LL0} and Theorem \ref{mainteoremGen}.

\begin{lemma} \label{LL0}
Let $C$ be a finite maximal code over
$A$ with $C \cap a^* = \{ a^p \}$,
for a prime number $p$.
Let $(P, S)$ be a factorization of $C$ such that $S = S_0 + S_1$.
If $S_1$ is a nonnull polynomial
with coefficients $0,1$, then $P_0, S_0$ are also polynomials
with coefficients $0,1$.
\end{lemma}

\begdim
Let $S_1 = \sum_{h \in H} a^{M_h} ba^{h}$, where $H$ and $M_h$ are finite,
nonempty sets of nonnegative integers, for $h \in H$.
Set $P_1 = \sum_{k \in K} a^kba^{L_k} - \sum_{t \in T} a^tba^{L'_t}$,
where $K, T$ are finite sets of nonnegative integers, $L_k, L'_t$
are finite multisets of nonnegative integers, for $k \in K, t \in T$,
and
$\supp( \sum_{k \in K} a^kba^{L_k}) \cap \supp( \sum_{t \in T} a^tba^{L'_t}) = \emptyset$.
Suppose that $P_0, S_0$ are not polynomials with coefficients $0,1$.
By Eq.~(\ref{eqCodes0}), we have $P_0 = -1$, $S_0 = -(1 + a + \ldots + a^{p-1})$
or  $P_0 = -(1 + a + \ldots + a^{p-1})$, $S_0 = -1$.

Assume that the first case holds.
Of course, $\C_1 \geq 0$ and, by Eqs.~(\ref{eqCodesR}), we have
\begin{eqnarray} \label{eqCodes1}
\C_1 &=& ba^{\{0, 1, \ldots , p-1\}} -  \sum_{h \in H} a^{M_h}(a-1) ba^{h}
+ \sum_{k \in K} a^kba^{L_k}(1 - a^p) + \sum_{t \in T} a^tba^{L'_t}(a^p - 1).
\end{eqnarray}
Let $m = \max \{ m' ~|~ m' \in M_{h'}, ~ h' \in H \}$.
Thus, for $h \in H$ such that $m = \max M_h$, we have
$( ba^{\{0, 1, \ldots , p-1\}} -  \sum_{h \in H} a^{M_h}(a-1) ba^{h}, a^{m+1}ba^h) < 0$.
By Eq.~(\ref{eqCodes1}) this implies
$$(\sum_{k \in K} a^kba^{L_k}(1 - a^p) + \sum_{t \in T} a^tba^{L'_t}(a^p - 1),
a^{m+1}ba^h) > 0.$$

Assume $(\sum_{k \in K} a^kba^{L_k}(1 - a^p), a^{m+1}ba^h) > 0$.
Hence, $m+1 \in K$ and $L_{m+1} \not = \emptyset$.
Let $\ell = \max L_{m+1}$. We now prove that $(\C_1, a^{m+1}ba^{\ell+p}) < 0$,
in contradiction with $\C_1 \geq 0$.
By the definition of $m$, we have
$(ba^{\{0, 1, \ldots , p-1\}} -  \sum_{h \in H} a^{M_h}(a-1) ba^{h}, a^{m+1}ba^{\ell+p}) \leq 0$
and by the definition of $\ell$, we have
$(\sum_{k \in K} a^kba^{L_k}(1 - a^p), a^{m+1}ba^{\ell + p}) < 0$.
If we had $(\sum_{t \in T} a^tba^{L'_t}(a^p - 1), a^{m+1}ba^{\ell + p}) > 0$,
then we would have $(\sum_{t \in T} a^tba^{L'_t}a^p, a^{m+1}ba^{\ell + p})
\geq  (\sum_{t \in T} a^tba^{L'_t}(a^p -1), a^{m+1}ba^{\ell+p}) > 0$, hence,
$m+1 \in T$, $\ell \in L'_{m+1}$ and
$\supp( \sum_{k \in K} a^kba^{L_k}) \cap \supp( \sum_{t \in T} a^tba^{L'_t})
\not = \emptyset$, a contradiction.

Therefore, $(\sum_{t \in T} a^tba^{L'_t}(a^p - 1), a^{m+1}ba^{h}) > 0$.
Hence, $m+1 \in T$ and $L'_{m+1} \not = \emptyset$.
Let $\ell = \min L'_{m+1}$. Then
$(ba^{\{0, 1, \ldots , p-1\}} -  \sum_{h \in H} a^{M_h}(a-1) ba^{h}, a^{m+1}ba^{\ell}) \leq 0$
by the definition of $m$, and
$(\sum_{t \in T} a^tba^{L'_t}(a^p - 1), a^{m+1}ba^{\ell}) < 0$
by the definition of $\ell$.
By Eq.~(\ref{eqCodes1}), this implies
$(\sum_{k \in K} a^kba^{L_k}(1 - a^p), a^{m+1}ba^{\ell}) > 0$
which yields $m+1 \in K$, $\ell \in L_{m+1}$ and
$\supp( \sum_{k \in K} a^kba^{L_k}) \cap \supp( \sum_{t \in T} a^tba^{L'_t})
\not = \emptyset$, again a contradiction.

Finally, assume $P_0 = -(1 + a + \ldots + a^{p-1})$, $S_0 = -1$.
Now $\C_1$ is defined by
\begin{eqnarray} \label{eqCodes1bis}
\C_1 &=& a^{\{0, 1, \ldots , p-1\}}b -  \sum_{h \in H} a^{M_h}(a^p-1) ba^{h}
+ \sum_{k \in K} a^kba^{L_k}(1 - a) + \sum_{t \in T} a^tba^{L'_t}(a -1).
\end{eqnarray}

Arguing as before, we consider $a^{m+p}ba^{h}$ with
$m = \max \{ m' ~|~ m' \in M_{h'}, ~ h' \in H \} = \max M_h$.
Since $(a^{\{0, 1, \ldots , p-1\}}b -  \sum_{h \in H} a^{M_h}(a^p-1) ba^{h}, a^{m+p}ba^{h}) < 0$
and $\C_1 \geq 0$, by Eq.~(\ref{eqCodes1bis}) we have
$(\sum_{k \in K} a^kba^{L_k}(1 - a) + \sum_{t \in T} a^tba^{L'_t}(a - 1), a^{m+p}ba^{h}) > 0$.

If $(\sum_{k \in K} a^kba^{L_k}(1 - a), a^{m+p}ba^{h}) > 0$, then
$m+p \in K$ and $L_{m+p} \not = \emptyset$.
Hence, for $\ell = \max L_{m+p}$ we have
$(\sum_{k \in K} a^kba^{L_k}(1 - a), a^{m+p}ba^{\ell + 1}) < 0$. Moreover,
$(a^{\{0, 1, \ldots , p-1\}}b -  \sum_{h \in H} a^{M_h}(a^p-1) ba^{h}, a^{m+p}ba^{\ell + 1}) \leq 0$
by the definition of $m$, and $(\sum_{t \in T} a^tba^{L'_t}(a - 1), a^{m+p}ba^{\ell + 1}) \leq 0$, since
otherwise
$a^{m+p}ba^{\ell} \in \supp( \sum_{k \in K} a^kba^{L_k}) \cap \supp( \sum_{t \in T} a^tba^{L'_t})$.
Thus $(\C_1, a^{m+p}ba^{\ell + 1}) < 0$, a contradiction.
If $(\sum_{t \in T} a^tba^{L'_t}(a -1), a^{m+p}ba^{h}) > 0$, then a similar
argument applies, and
for $\ell = \min L'_{m+p}$, we get $(\C_1, a^{m+p}ba^{\ell}) < 0$, again a contradiction.
\enddim

The following lemma is needed for the proof of
Theorem \ref{mainteoremGen}.

\begin{lemma} \label{LL1}
Let $S$ be a polynomial in
$\N \langle A \rangle$
such that if the word $wba^j$ is in $\supp(S)$, then $a^j$ is also
in $\supp(S)$.
Let $S_0 = a^J$ with
$J = \{0 \}$ or $J = \{0, 1, \ldots , p-1 \}$,
where $p$ is a nonnegative number.
Let $r \geq 0$ and
assume that $P_{0}, P_{1}, \ldots , P_{r+1} \in \Z \langle A \rangle$
are polynomials such that
$$\sum_{i+h=r}P_{i}bS_{h} +
\sum_{i+h=r + 1}P_{i}(a-1)S_{h}  \geq 0.$$
If $P_{0}, P_1, \ldots , P_{r} \in \N \langle A \rangle$,
then $P_{r+1} \in \N \langle A \rangle$.
\end{lemma}

\begdim
Assume that $J = \{0, 1, \ldots , p-1 \}$ (if $J = \{0 \}$
we may apply the following argument with $p = 1$).
By contradiction, let
$P_{r+1} = P'_{r+1} - P''_{r+1}$ with
$P'_{r+1}, P''_{r+1} \in \N \langle A \rangle$,
$\supp(P'_{r+1}) \cap \supp(P''_{r+1}) = \emptyset$
and $P''_{r+1} \not = 0$.
Let $\ell = \max \{\ell' ~|~ \exists x \in A^* ~
xba^{\ell'} \in \supp(P''_{r+1}) \}$
and let $x$ be a word such that
$y = xba^{\ell} \in \supp(P''_{r+1})$.

\noindent
By hypothesis, for any word $wba^j \in \supp(S)$, the nonnegative integer
$j$ is less than $p$.
Hence,
$(\sum_{i+h=r}P_{i}bS_{h} + \sum_{i+h=r + 1, \; h \not = 0}P_{i}(a-1)S_{h}, ya^p) = 0$.
By the definition of $y$, we also have
$(P''_{r+1}(a^p-1),  y a^p) = (P''_{r+1}a^p, ya^p) > 0$.
Thus
\begin{eqnarray*}
0 &\leq& (\sum_{i+h=r}P_{i}bS_{h} + \sum_{i+h=r + 1}P_{i}(a-1)S_{h}, ya^p) \\
&= &
(P_{r+1}(a-1)a^J, ya^p) = (P_{r+1}(a^p-1) , y a^p) \\
& = &
(P'_{r+1}(a^p-1),  y a^p) -
(P''_{r+1}(a^p-1),  y a^p) \\
& \leq &
(P'_{r+1}a^p, ya^p) - (P''_{r+1}(a^p-1),  y a^p) \\
& < &
(P'_{r+1}a^p, ya^p) = (P'_{r+1}, y).
\end{eqnarray*}
In conclusion, $(P'_{r+1}, y) > 0$ and
$ y \in \supp(P'_{r+1}) \cap \supp(P''_{r+1})$,
a contradiction.
\enddim

\begin{theorem} \label{mainteoremGen}
Let $C$ be a finite maximal code such that $a^p \in C$, for a
prime number $p$. Let
$P \in \Z \langle A \rangle$ and let
$S$ be a polynomial in
$\N \langle A \rangle$ such that
if the word $wba^j$ is in $\supp(S)$ then $a^j$ is also
in $\supp(S)$.
If $(P,S)$ is a factorization
for $C$, then $P,S$ have coefficients $0,1$.
\end{theorem}

\begdim
Let $P = P_0  + P_1 + \ldots + P_k$ and let
$S$ be a polynomial in
$\N \langle A \rangle$
such that if the word $wba^j$ is in $\supp(S)$ then $a^j$ is also
in $\supp(S)$.
Assume that $(P,S)$ is a factorization
for a finite maximal code $C$ and $a^p \in C$, where $p$ is a
prime number.
Thus $P_0$ has coefficients $0,1$ and
$\C_{r+1}$ is defined by Eqs.~(\ref{eqCodesR}), for any $r \geq 0$.
By using induction and Lemma \ref{LL1},
we can prove that $P_0, P_1,  \ldots , P_k \in \N \langle A \rangle$.
Hence $P \in \N \langle A \rangle$ and,
by Theorem \ref{ine}, $P, S$ have coefficients
$0, 1$.
\enddim

\begin{remark}
{\rm Lemma \ref{LL1} is no longer true if we drop
the hypothesis that $wba^j \in \supp(S)$ only if $a^j \in \supp(S)$,
even if $S$ is a polynomial with
coefficients $0,1$. Indeed, let $s,t, n \in \N$, with $s \geq 1$,
$t \geq 0$, $n \geq 1$, let $P_0 = 1$,
$P_1 = ba^{\{0, 1, \ldots, t \}n} - a^sba^{tn}$,
$S = a^{\{0, 1, \ldots, n-1\}} +  a^{\{0, 1, \ldots , s-1\}}ba^{(t+1)n}$.
Then
$P_0bS_0 + P_1(a-1)S_0 + P_0(a-1)S_1 = b a^{\{1, \ldots , n-1 \}} + a^sba^{tn}$
is a polynomial with coefficients $0,1$.
However, we do not know whether Theorem \ref{mainteoremGen} is still true
without the aforementioned hypothesis.
As already stated in Section \ref{introduction},
in \cite{KH} the authors formulated the following conjecture:
if $(P,S)$ is a factorization
for a finite maximal code over a one-letter alphabet and $S$
has coefficients $0,1$, then $(P,S)$ is positive.
Notice that Theorem \ref{mainteoremGen}
is connected with a generalization
of this conjecture to alphabets with size greater
than one.}
\end{remark}

\begin{theorem} \label{mainteorem1}
Let $C$ be a finite maximal code such that $a^p \in C$, for a
prime number $p$. Let
$S = S_0 + S_1$ be a polynomial such
that if the word $a^iba^j$ is in $\supp(S)$ then the word $a^j$ is also
in $\supp(S)$.
If $(P,S)$ is a factorization
for $C$, then $(P,S)$ is positive.
\end{theorem}

\begdim
Assume that $C$ and $S$ are as in the statement.
Let $(P,S)$ be a factorization
for $C$. If $S_1 = 0$ then by Theorem \ref{P(a)}
$(P,S)$ is positive. Otherwise, by
Theorem \ref{T2}, either $S_1$ or $-S_1$ has coefficients
$0,1$. In the first case, by Lemma \ref{LL0},
$P_0, S_0$ have coefficients $0,1$.
Thus, Theorem \ref{mainteoremGen} applies
to $(P,S)$ and $P,S$ have coefficients $0,1$.
In the second case, arguing as before on the factorization
$(-P, -S)$ for $C$, we can prove that $-P,-S$ have coefficients $0,1$.
Hence, in both cases $(P,S)$ is positive.
\enddim

\subsection{$(p,4)$-codes are positively factorizing} \label{main2}

In this section, we consider $(p,4)$-codes,
i.e., $4$-codes $C$ such that
$C \cap a^* = \{ a^p \}$, for a prime number $p$.
We show that they are positively factorizing and
have only positive factorizations.
Looking at Eqs.~(\ref{eqCodesR}),
factorizations for a $4$-code may be divided into two sets,
as described in Lemma \ref{L0}.

\begin{lemma} \label{L0}
Let $C$ be a $4$-code, let
$(U, V)$ be a factorization for $C$.
Then for $(P,S) = (U, V)$ or for $(P,S) = (V^{\sim}, U^{\sim})$
one of the following two conditions
is satisfied.
\begin{itemize}
\item[(1)]
$P = P_0 + P_1 + P_2 +P_3$,
$S = S_0$, with $P_3 \not = 0$, $S_0 \not = 0$.
\item[(2)]
$P = P_0 + P_1 + P_2$, $S = S_0 + S_1$, with $P_2 \not = 0$, $S_1 \not = 0$.
\end{itemize}
\end{lemma}

\begin{theorem} \label{mainteorem2}
Let $p$ be a prime number.
Any $(p,4)$-code $C$ is
positively factorizing.
Moreover, each factorization for $C$ is
a positive factorization for $C$.
\end{theorem}

\begdim
Let $p$ be a prime number, let $C$ be a $(p,4)$-code
and let $(U,V)$ be a factorization for $C$.
By Lemma \ref{L0}, $(P,S) = (U,V)$
or $(P,S) = (V^{\sim}, U^{\sim})$,
satisfies item $(1)$ or item $(2)$ in this
lemma.
If $(P,S) = (U,V)$
satisfies item $(1)$ in Lemma \ref{L0},
then $(P,S)$ is positive
by Theorem \ref{P(a)}.
Assume that $(P,S) = (U,V)$
satisfies item $(2)$ in Lemma \ref{L0}.
By Theorem \ref{T2}, $(P,S)$
also satisfies one of the following
four conditions:
\begin{itemize}
\item[(i)]
$P_2, S_1$ have coefficients $0,1$ and $P_1 \in \N \langle A \rangle$.
\item[(ii)]
$P_2, S_1$ have coefficients $0,1$ and if the word
$a^iba^j$ is in $\supp(S)$, then the word $a^j$ is also
in $\supp(S)$.
\item[(iii)]
$-P_2, -S_1$ have coefficients $0,1$ and $-P_1 \in \N \langle A \rangle$.
\item[(ii)]
$-P_2, -S_1$ have coefficients $0,1$ and if the word
$a^iba^j$ is in $\supp(S)$, then the word $a^j$ is also
in $\supp(S)$.
\end{itemize}

In view of Lemma \ref{LL0} and Theorem \ref{ine}, it is obvious that
if $(P,S)$ satisfies
item (i), then
$P,S$ have coefficients $0,1$.
By Theorem \ref{mainteorem1}, it is also clear that
if $(P,S)$ satisfies
item (ii), then
$P,S$ have coefficients $0,1$. Finally,
if $(P,S)$ satisfies
item (iii) or (iv), then
$-P, -S$ have coefficients $0,1$ and $(P,S)$ is positive.

Assume now that $(P,S) = (V^{\sim}, U^{\sim})$
satisfies item (1) or item (2) in Lemma \ref{L0}.
The pair $(V^{\sim}, U^{\sim})$ is a factorization
for the $(p,4)$-code $C^{\sim}$ and the above arguments
prove that $(V^{\sim}, U^{\sim})$ is positive. Hence
$(U,V)$ is also positive.
\enddim

\subsection{Partially positive factorizations} \label{main3}

We end this section with another result:
if $(P,S)$ is a factorization for $C$ such that
$P_0 = a^I$, $S_0 = a^J$ and $S_1 = \sum_{j \in J} a^{M_{j}}b a^j$,
where $I, J, M_j $, for $j \in J$, are finite subsets of $\N$, then
$P_1$ has nonnegative coefficients (Theorem \ref{mainteorem3}).
Let $(P,S)$ be a factorization for $C$ satisfying the above conditions. Set
$P_1 =  \sum_{i \in I'}a^ib(a^{L_i} - a^{L'_i})$, where
$I'$ is a finite subset of $\N$
and $a^{L_i}, a^{L'_i }$
are polynomials in $\N[a]$ such that
$L_i \cap L'_i = \emptyset$,
for each $i \in I'$.
By Eqs.~(\ref{eqCodesR}) for $r = 0$, we have
\begin{eqnarray} \label{EqCodes1}
a^Iba^J +
\sum_{i \in I'}a^ib(a^{L_i} - a^{L'_i})(a-1)a^J +
\sum_{j \in J}a^I(a-1)a^{M_j}ba^j \geq 0.
\end{eqnarray}
Notice that $P_0(a-1)S_0 = a^n -1$,
i.e., $a^I a^J = 1 + a + \ldots + a^{n-1}$, where $a^n \in C$.
These pairs $(I,J)$ of subsets
of $\N$ can be constructed by a
method given in \cite{KRR}
and here they will be called {\it Krasner factorizations}
of the finite cyclic group $\Z_{n}$ of order $n$.
Theorem \ref{mainteorem3} will be proved through
Lemmas \ref{L2} -- \ref{L4C}.
These lemmas will also be used in Sections \ref{construction1}
and \ref{construction4}.

\begin{lemma} \label{L2}
Let $(I,J)$ be a Krasner factorization
of $\Z_{n}$. Let $I'$, $M_{j}$, with $j \in J$,
be finite subsets of $\N$,
let $a^{L_i}, a^{L'_i} \in  \N[a]$,
with $i \in I'$. Assume that Eq.~(\ref{EqCodes1}) holds.
Then, for each $i \in I'$, there
exists $k_i \in \N$ such that
\begin{eqnarray*}
a^{L_i}(a-1)a^J - a^{L'_i}(a-1)a^J
+ k_ia^J &\geq& 0.
\end{eqnarray*}
\end{lemma}

\begdim
Assume that $I'$, $M_{j}$, with $j \in J$,
$a^{L_i}$, $a^{L'_i}$, with $i \in I'$, are as in the
statement. By using Eq.~(\ref{EqCodes1}), we have:
\begin{eqnarray*}
& & \forall i \in I' \quad 0 \leq (a^I, a^i)a^J +
(a^{L_i} - a^{L'_i})(a-1)a^J +
\sum_{j \in J}(a^I(a-1)a^{M_j}, a^i) a^j \\
& & \leq (a^I, a^i)a^J +
(a^{L_i} - a^{L'_i})(a-1)a^J +
\sum_{j \in J, \; (a^I(a-1)a^{M_j}, a^i) \geq 0}
(a^I(a-1)a^{M_j}, a^i) a^j \\
& & \leq a^{L_i}(a-1)a^J - a^{L'_i}(a-1)a^J
+ k_ia^J,
\end{eqnarray*}
where $k_i = k'_i + (a^I, a^i)$,
$k'_i = \max \Gamma_i$, $\Gamma_i = \{0 \} \cup
\{\gamma_j ~|~ \gamma_j = (a^I(a-1)a^{M_j}, a^i) \geq 0, j \in J \}$.
\enddim

\begin{lemma} \label{L3}
Let $k \in \N$. Let
$a^X, a^{X'} \in \N[a]$,
with $X \cap X' =  \emptyset$.
If we have
\begin{eqnarray} \label{Eq3}
a^{X}(a-1) - a^{X'}(a-1) + k & \geq & 0,
\end{eqnarray}
then $a^{X'} = 0$. Furthermore, if $a^X$ is a nonnull polynomial
then $k > 0$.
\end{lemma}

\begdim
By contradiction, assume that
$a^{X'} \not = 0$ and let
$x = \max \{x' ~|~ x' \in X' \}$.
Thus we have $x + 1 > 0$ and
$(- a^{X'}(a-1), a^{x+1}) < 0$.
Hence, in view of Eq.~(\ref{Eq3}), we have
\begin{eqnarray*}
0 & \leq & (a^{X}(a-1) - a^{X'}(a-1) + k, a^{x+1}) \\
  & = & (a^{X}(a-1) - a^{X'}(a-1), a^{x+1}) \\
  & < & (a^{X}(a-1), a^{x+1}) \\
& \leq & (a^{X}a, a^{x+1}),
\end{eqnarray*}
which yields $x \in X \cap X'$,
a contradiction.
Thus $a^{X'} = 0$. If $a^X \not = 0$,
let $x = \min \{x' ~|~ x' \in X \}$. Since
$(a^{X}(a-1), a^x) < 0$, we have $k > 0$.
\enddim

Let $a^H \in \N[a]$, $n \in \N$
and $t \in \{0, \ldots , n-1 \}$.
We set $[H]_t= \{ h \in H ~|~ h = t \! \pmod{n} \}$.
Notice that $[H]_t$ could be a multiset:
any element $h \in H$
such that $h = t \! \pmod{n}$ is in $[H]_t$
with the same multiplicity
as in $H$.

\begin{lemma} \label{L4}
Let $(I,J)$ be a Krasner factorization
of $\Z_{n}$, let $k \in \N$.
Let $a^L, a^{L'} \in \N[a]$,
with $L \cap L' = \emptyset$.
If we have
\begin{eqnarray} \label{Eq4}
a^{L}(a-1)a^Ja^I - a^{L'}(a-1)a^Ja^I + ka^Ja^I & \geq & 0,
\end{eqnarray}
then $a^{L'} = 0$. Furthermore, if $a^L$
is a nonnull polynomial
then $k > 0$.
\end{lemma}

\begdim
By Eq.~(\ref{Eq4}), since $(I,J)$ is
a Krasner factorization
of $\Z_{n}$, we have:
\begin{eqnarray*}
0 & \leq & a^{L}(a-1)a^Ja^I -
a^{L'}(a-1)a^Ja^I + ka^Ja^I \\
& = & a^{L}(a^n -1) - a^{L'}(a^n-1)
+ k \frac{a^n -1}{a -1},
\end{eqnarray*}
which yields:
\begin{eqnarray} \label{Eq4bis}
\forall t \in \{0, \ldots , n-1\}, & &
a^{[L]_t}(a^n -1) - a^{[L']_t}(a^n-1) + k a^t \geq 0.
\end{eqnarray}
By erasing $a^t$ and by changing $a^n$ with $a$ in
each term of this inequality, we get an inequality
as in Eq.~(\ref{Eq3}). Precisely, for all
$t \in \{0, \ldots , n-1\}$, let
$a^{X_t}, a^{X'_t} \in \N[a]$ be defined as follows:
\begin{eqnarray*}
a^{X_t} &=& \sum_{x \in \N} (a^{[L]_t}, a^{t + x n})a^x, \quad
a^{X'_t} = \sum_{x' \in \N} (a^{[L']_t}, a^{t + x' n})a^{x'}.
\end{eqnarray*}
As a direct consequence we have:
\begin{eqnarray*}
& &
X_t \not = \emptyset \Leftrightarrow
[L]_t \not = \emptyset, \quad
X'_t \not = \emptyset \Leftrightarrow
[L']_t \not = \emptyset.
\end{eqnarray*}
Furthermore, for any $x \in \N$, we also have
$$(a^{[L]_t}(a^n -1) - a^{[L']_t}(a^n-1) + k a^t, a^{xn + t})
= (a^{X_t}(a-1) - a^{X'_t}(a-1) + k, a^x).$$
Since for any $y, t \in \N$ we have
$(a^{[L]_t}(a^n -1) - a^{[L']_t}(a^n-1) + k a^t, a^y) \not = 0$
if and only if $y = xn +t$, the above relation and Eq.~(\ref{Eq4bis})
show that $a^{X_t}(a-1) - a^{X'_t}(a-1) + k \geq 0$.
Thus, in view of Lemma \ref{L3},
$a^{X'_t} = 0$, for all $t$. The latter relation yields
$a^{[L']_t} = 0$ for all $t$, i.e.,
$a^{L'} = 0$.
Finally, if $a^L \not = 0$ then there is $t$ such that
$[L]_t \not = \emptyset$. Consequently,
$a^{X_t} \not = 0$ and $k > 0$, once again
by Lemma \ref{L3}.
\enddim

\begin{lemma} \label{L4C}
Let $(I,J)$ be a Krasner factorization
of $\Z_{n}$, let $k \in \N$.
Let $a^L, a^{L'} \in \N[a]$,
with $L \cap L' = \emptyset$.
If we have
\begin{eqnarray} \label{Eq4C}
a^{L}(a-1)a^J - a^{L'}(a-1)a^J + ka^J & \geq & 0,
\end{eqnarray}
then $a^{L'} = 0$. Furthermore, if $a^L$
is a nonnull polynomial
then $k > 0$.
\end{lemma}

\begdim
By Eq.~(\ref{Eq4C}) and since $a^I \geq 0$, we have:
\begin{eqnarray*}
& & a^{L}(a-1)a^Ja^I -
a^{L'}(a-1)a^Ja^I + ka^Ja^I \geq 0.
\end{eqnarray*}
Thus the conclusion follows by Lemma \ref{L4}.
\enddim

\noindent
We have proved the following result.

\begin{theorem} \label{mainteorem3}
Let $(P,S)$ be a factorization for $C$ such that
$P_0 = a^I$, $S_0 = a^J$ and $S_1 = \sum_{j \in J} a^{M_{j}}b a^j$,
where $(I,J)$ is a Krasner factorization
of $\Z_{n}$ and
$M_j $ is a finite subset of $\N$, for $j \in J$.
Then $P_1$ has nonnegative coefficients.
\end{theorem}

\begdim
Let $(P,S)$ be a factorization for $C$ with $P_0, S_0, S_1$ as in the statement.
Let $I'$ be a finite subset of $\N$
and let $a^{L_i}, a^{L'_i }$ be
polynomials in $\N[a]$ such that
$P_1 = \sum_{i \in I'}a^ib(a^{L_i} - a^{L'_i})$, where
$L_i \cap L'_i = \emptyset$,
for $i \in I'$.
Thus $(P,S)$ satisfies
Eq.~(\ref{EqCodes1}).
By Lemma \ref{L2}, for each $i \in I'$ there
exists $k_i \in \N$ such that
$a^{L_i}(a-1)a^J - a^{L'_i}(a-1)a^J
+ k_ia^J \geq 0$.
Thus, $a^{L'_i} = 0$
for each $i \in I'$ (Lemma \ref{L4C}) and $P_1 \in  \N \langle A \rangle$.
\enddim

\section{Positively
factorizing codes} \label{outline1}

As observed in \cite{BPR}, the aim of the theory of codes
is to give a structural description of codes in a way that
allows their construction. This has not yet been accomplished,
except for some special families of codes.
In particular, a still open problem is a structural description
of positively factorizing codes.
Once again, this description has been achieved for particular
classes of codes, through the construction of their factorizations.
In this paper, we extend this construction to a larger class,
as outlined in Section \ref{outlinePFC}.
The obtained results once again show relations
between factorizations of finite maximal codes
and factorizations of finite cyclic groups, whose definition
is recalled in Section \ref{FCG}.

\subsection{Factorizations of cyclic groups} \label{FCG}

A pair $(T,R)$ of subsets of $\N$
is a factorization of $\Z_{n}$ if,
for each $i$ in $\{0, \ldots, n-1 \}$, there
exists a unique pair $(t,r) \in T \times R$ such that
$i = t + r \pmod{n}$. We are interested in a special
class of factorizations, defined in \cite{HA50}
and called {\it Haj\'os factorizations} here.
There are relations between the structure
of positively factorizing codes and Haj\'os factorizations
of $\Z_{n}$.
These relations have been highlighted by
a characterization of Haj\'os factorizations given in \cite{DF96}:
$(T,R)$ is a Haj\'os factorization
of $\Z_{n}$ if and only if there is a Krasner factorization
$(I,J)$ of $\Z_{n}$ and a pair $(M,L)$ of
finite subsets of $\N$ such that
$a^T = a^M(a-1)a^I + a^I \geq 0$,
$a^R = a^L(a-1)a^J + a^J \geq 0$.
The structure of the above subsets $M,L$, and therefore of the pairs
$(T, R)$, has been described in \cite{DF89}.
As a consequence, the following result has been stated
(see Remark 6.7 in \cite{DF89}).

\begin{proposition} \label{Ul}
Let $(I,J)$ be a Krasner factorization of $\Z_{n}$,
let $M$ be a finite subset of $\N$ and let
$a^T = a^{M}(a-1)a^{I} + a^{I}$.
If $a^T$ has nonnegative coefficients,
then $a^T$ has coefficients $0,1$.
\end{proposition}

As stated in Proposition \ref{holes},
subsets $M,L$ satisfy another equation too.
We also need the following known result
(Lemma 3.2 (ii), (iii), (iv) in \cite{DF93}).

\begin{lemma} \label{LG0}
Let $k, n \in \N$.
\begin{itemize}
\item[(i)]
If $H$ is a finite subset of $\N$ and $(a^{H}(a-1) + k)(a^n -1)/(a -1) \geq 0$,
then the polynomial
$(a^{H}(a-1) + 1)(a^n -1)/(a -1)$ has coefficients $0,1$.
\item[(ii)]
If $a^H \in \N[a]$ and
$(a^{H}(a-1) +1)(a^n -1)/(a -1) \geq 0$,
then $a^H$ has coefficients $0,1$.
\end{itemize}
\end{lemma}

\begin{proposition} \label{holes}
Let $(I,J)$ be a Krasner factorization of $\Z_{n}$,
let $M, L$ be finite subsets of $\N$ such that
$$a^T = a^M(a-1)a^I + a^I \geq 0, \quad
a^R = a^L(a-1)a^J + a^J \geq 0.$$
Then $a^M(a-1)a^L + a^M + a^L$ is a polynomial with
coefficients $0,1$.
\end{proposition}

\begdim
Let $a^H, a^{H'} \in \N[a]$,
with $H \cap H' = \emptyset$,
and assume
$a^M(a-1)a^L + a^M + a^L = a^H - a^{H'}$.
An easy computation shows that
$$a^T a^R= a^{H}(a-1)a^Ja^I - a^{H'}(a-1)a^Ja^I + a^Ja^I \geq 0.$$
Thus $a^{H'} = 0$, by Lemma \ref{L4}.

Then $(a^{H}(a-1)+1)(a^n -1)/(a -1) \geq 0$
and $a^H = a^M(a-1)a^L + a^M + a^L$ has coefficients $0,1$
by Lemma \ref{LG0} (ii).
\enddim

We will see that our construction of positive factorizations is
strongly related to
Haj\'os factorizations $(T,R)$ satisfying an additional
hypothesis: for the corresponding pair $(M,L)$
we have
$a^M(a-1)a^L + a^L \geq 0$ or $a^M(a-1)a^L + a^M \geq 0$.
In this case we say that $(T,R)$ is a
{\it strong Haj\'os factorization}.

In \cite{LAM97}, the author gave a construction
of an infinite family of
Haj\'{o}s factorizations $(T, R)$ of $\Z_{n}$ which
are not strong, i.e., which are
such that, for the corresponding pair $(M,L)$, neither
$a^M(a-1)a^L + a^L$ nor $a^M(a-1)a^L + a^M$ is a polynomial
with nonnegative coefficients.
The non-strong Haj\'{o}s factorization
$(T,R)=(\{0,4,8,12,16,20\}, \{0,3,6,21 \})$
of $\Z_{24}$ is an element
of this family and the corresponding
pair $(M,L)$
is $(\{2,3\}, \{1,9,11,13\})$.

\subsection{Outline of the results on positively factorizing codes}
\label{outlinePFC}

So far, factorizations $(P,S)$ for finite maximal
codes $C$ have been constructed for $3$-codes
or when $\supp(P)$ or
$\supp(S)$ is a subset of $a^*$. Factorizations for $1$- and $2$-codes
belong to the latter family.
As said, all these factorizations are positive, i.e.,
$P,S \in \N \langle A \rangle$ or
$-P, -S \in \N \langle A \rangle$. In the
description of their structure, as well as for
other positive factorizations, we will assume
$P,S \in \N \langle A \rangle$.

Let $P,S \in \N \langle A \rangle$.
Then $(P,S)$ is a positive factorization for a $1$-code
if and only if $P = a^I$ and $S = a^J$,
for a Krasner factorization $(I,J)$ of $\Z_{n}$ \cite{RES}.
Starting with these pairs $(a^I, a^J)$, one may inductively
construct all positive factorizations $(P,S)$ with $P$ or $S$ in $\N[a]$
as follows \cite{DF89}.
Assume $P \in \N[a]$.
We have $P = a^I$, $S = S_0 + S_1 + \ldots + S_t$, with
$S_0 = a^J$ and there are finite subsets $M_w$
of $\N$ such that $S_i = \sum_{w \in \supp(S_{i-1})}a^{M_w}bw$, $1 \leq i \leq t$,
with $a^{T_w}= a^{M_w}(a-1)a^I + a^I \geq 0$.
Therefore $(T_w, J)$ is a strong Haj\'{o}s factorization of $\Z_{n}$
since for the corresponding pair $(M_w, \emptyset)$, we have
$a^{M_w} \geq 0$.
Of course, if $S \in \N[a]$, then
$(S^{\sim}, P^{\sim})$ is as above.

In this paper,
starting with the ``simplest'' pairs $(a^I, a^J + \sum_{j \in J}a^{M_j}ba^j)$
in the above family,
we give a recursive construction of all
positive factorizations $(P,S)$ for a maximal code $C$ with
$S = a^J + \sum_{j \in J}a^{M_j}ba^j$. Notice that when $n$ is a prime number, the
factorizations $(P,S)$
are those mentioned in Theorem \ref{mainteorem1}.
We begin with a characterization of the words $C_1$
with one occurrence of $b$ in Section \ref{construction1}.
Then, in Section \ref{constructionC},
we prove that each $m$-code $C$ having a factorization $(P,S)$,
with $S$ as above, may be obtained
from an $(m-1)$-code $C'$ having a factorization $(P', S)$.

Positive factorizations $(P,S)$ with $\supp(P) \subseteq a^* \cup a^*ba^*$ and
$S = a^J + \sum_{j \in J}a^{M_j}ba^j$ have already been characterized in \cite{DF93}.
As a matter of fact, if $(U,V)$ is a positive factorization
for a $3$-code such that neither $U$ nor $V$ is
in $\N[a]$, then for $(P,S) = (U,V)$ or $(P,S) = (V^{\sim}, U^{\sim})$,
we have
$P = a^I + \sum_{i \in I'}a^iba^{L_i}$, $S = a^J + \sum_{j \in J}a^{M_j}ba^j$,
where $I'$, $L_i$, for $i \in I'$, and $M_j$, for $j \in J$, are finite subsets of $\N$
such that
\begin{eqnarray} \label{3.1}
\forall j \in J & & a^{T_j} = a^{M_{j}}(a-1)a^{I} + a^{I} \geq 0,
\end{eqnarray}
\begin{eqnarray} \label{3.0}
& & \{i \in I' ~|~ L_i \not = \emptyset \} \subseteq \cup_{j \in J} T_j,
\end{eqnarray}
\begin{eqnarray} \label{3.2}
\forall i \in I'& & a^{R_{i}} = a^{L_{i}}(a-1)a^{J} + a^{J} \geq
a^{L_{i}}(a-1)a^{J} + a^{J_i} \geq 0,
\end{eqnarray}
where $J_i = \{j \in J ~|~ i \in T_j \}$,
\begin{eqnarray} \label{3.3}
\forall j \in J, i \in I' \setminus I & & a^{L_{i}}(a-1)a^{M_{j}}+
a^{L_{i}} \geq 0.
\end{eqnarray}
Moreover, there are $i \in I'$ and $j \in J$ such that $L_i \not = \emptyset$
and $M_j \not = \emptyset$.

\begin{remark} \label{0}
{\rm Of course, if $M_{j'} = \emptyset$, then $T_{j'} = I$
and Eqs.~(\ref{3.1}) and (\ref{3.3}) are satisfied for $j'$. Analogously,
if $L_{i'} = \emptyset$, then Eqs.~(\ref{3.2}) and (\ref{3.3}) are satisfied
for $i'$.}
\end{remark}

\begin{remark} \label{2}
{\rm Note that in the characterization of
the positive factorizations $(P,S)$ given in \cite{DF93}, it
is also required that
$a^{M_{j}}(a-1)a^{L_{i}}+ a^{M_{j}} +
a^{L_{i}} \geq 0$, for $i \in I' \cap I, j \in J$.
In view of Eqs.~(\ref{3.1}), (\ref{3.2}), Proposition \ref{holes}
applies to $L_i$, $M_j$ and shows that the above polynomial has
coefficients $0,1$. Therefore,
we may omit this condition.}
\end{remark}

Section \ref{construction4}
deals with some positive factorizations
for $4$-codes which are described below.
Lemma \ref{M0} is the counterpart of Lemma \ref{L0}
for positive factorizations.

\begin{lemma} \label{M0}
Let $C$ be a $4$-code, let
$(U,V)$ be a factorization for $C$, with $U, V \in \N \langle A \rangle$.
Then, for $(P,S) = (U,V)$ or $(P,S) = (V^{\sim}, U^{\sim})$,
one of the following three conditions is satisfied.
\begin{itemize}
\item[(1)]
$P = P_0 + P_1 + P_2 +P_3 \in \N \langle A \rangle$,
$S = S_0 \in \N[a]$, with $P_3 \not = 0$, $S_0 \not = 0$.
\item[(2)]
$P = a^I + \sum_{i \in I'}a^iba^{L_i} + \sum_{w \in X_1}wba^{L_w}$,
$S = a^J + \sum_{j \in J} a^{M_{j}}b a^j$, where
$(I,J)$ is a Krasner factorization of $\Z_{n}$,
$I', L_i, M_j, L_w$ are finite subsets of $\N$, for any
$i, j, w$, and $X_1$ is
a finite subset of $a^*ba^*$.
Moreover, $\sum_{w \in X_1}wba^{L_w} \not = 0$, $\sum_{j \in J} a^{M_{j}}b a^j \not = 0$.
\item[(3)]
$P = a^I + \sum_{i \in I'}a^iba^{L_i} +
\sum_{i \in I', \ell \in L_i}a^iba^{\ell}ba^{L_{i,\ell}}$,
$S = a^J + \sum_{j \in J'} a^{M_{j}}b a^j$, where
$(I,J)$ is a Krasner factorization of $\Z_{n}$,
$I', J', L_i, M_j, L_{i,\ell}$ are finite subsets of $\N$,
for any $i,j, \ell$. Moreover,
$\sum_{i \in I', \ell \in L_i}a^iba^{\ell}ba^{L_{i,\ell}} \not = 0$,
$\sum_{j \in J'} a^{M_{j}}b a^j \not = 0$.
\end{itemize}
\end{lemma}

\begdim
The statement is a direct consequence of Lemma \ref{L0}
and Theorem \ref{T2}.
\enddim

We have already described the structure
of the positive factorizations for $4$-codes satisfying
item (1) in Lemma \ref{M0}.
Furthermore, $4$-codes having positive factorizations that satisfy
item (2) in Lemma \ref{M0}
belong to the class considered in Section \ref{constructionC}.
Positive factorizations $(P,S)$,
satisfying item (3) in Lemma \ref{M0},
will be handled in Section \ref{construction4}
when $I' = I$, i.e., when all words $a^ibw$ in $\supp(P)$ are such
that $i \in I$.

Finally, let $\Omega(n)$ be the number of factors in the prime
factorization of $n \in \N$.
We recall that the structure of the words
in $C_1 = C \cap a^*ba^*$ has been investigated in
\cite{DF05}. A characterization of the words in $C_1$ has
been obtained when $a^n \in C$ with $\Omega(n) \leq 2$ \cite{DF07b}.

\section{Words in $C$ with
one $b$: a special case} \label{construction1}

Let $(P,S)$ be a
factorization for a finite maximal code $C$ with
$P, S \in \N \langle A \rangle$.
Set $P_0 = a^I$, $S_0 = a^J$, where $(I,J)$ is a Krasner
factorization of $\Z_{n}$.
Lemma \ref{LG1} characterizes polynomials $P_1, S_1$ under
the hypothesis $S_1 = \sum_{j \in J} a^{M_j} ba^j$
or $P_1 = \sum_{i \in I} a^i ba^{L_i}$.
Loosely speaking, this result states that the set
$C_1 = C \cap a^* ba^*$ of the words with one occurrence of $b$
is the same as in a $3$-code. The proof of Lemma \ref{LG1}
is the same as in \cite{DF93} and it is reported here for the sake
of completeness.
This result will also be used in Sections \ref{constructionC}
and \ref{construction4}.

\begin{proposition} \label{CO1}
Let $(I,J)$ be a Krasner factorization of $\Z_{n}$,
let $L$ be a finite subset of $\N$ and let $k$ be a positive integer such that
$a^{L}(a-1)a^{J} + k a^{J} \geq 0$.
Then, for any $j \in J$, we have
$(a^{L}(a-1)a^{J} + a^{J}, a^j) \leq 1$.
\end{proposition}

\begdim
Assume that there exists $k > 0$
such that $a^{L}(a-1)a^{J} + k a^{J} \geq 0$. Thus,
$$(a^{L}(a-1)a^{J} + k a^{J})a^I = (a^{L}(a -1) + k)(a^n -1)/(a -1)  \geq 0.$$
Therefore, by Lemma \ref{LG0} (i),
$(a^{L}(a -1) + 1)(a^n -1)/(a -1)$ is a polynomial
with coefficients $0,1$.
Assume that
$(a^{L}(a-1)a^J + a^J, a^j) \geq 2$ with $j \in J$.
Notice that
$$(a^{L}(a -1) + 1)(a^n -1)/(a -1)= a^{L}(a-1)a^Ja^I + a^Ja^I = a^{L}(a-1)a^J + a^J +
(a^{L}(a-1) + 1)a^Ja^{I \setminus 0}.$$
Thus, we obtain
$((a^{L}(a-1) + 1)a^Ja^{I \setminus 0}, a^j) < 0$, i.e.,
$$\exists q \in \N, ~ i \in I \setminus 0 ~:~ q + i = j,
\; (a^{L}(a-1)a^J + a^J, a^q) < 0.$$
On the other hand, since
$a^{L}(a-1)a^{J} + k a^{J} \geq 0$, we have $k > 1$ and
$q \in J$. This is impossible since $(I,J)$ is a Krasner factorization of $\Z_{n}$
and $i + q = 0 + j$ with
$i \in I \setminus 0$, $0 \in I$, $q,j \in J$.
\enddim

\begin{lemma} \label{LG1}
Let $(I,J)$ be a Krasner factorization of $\Z_{n}$,
let $I', L_i, M_j$ be finite subsets of $\N$.
We have
$$\C_1 = a^Iba^J + \sum_{i \in I'}a^iba^{L_i}(a-1)a^J +
\sum_{j \in J} a^I(a-1)a^{M_j}ba^j \geq 0$$
if and only if $I', L_i, M_j$  satisfy Eqs.~(\ref{3.1})--(\ref{3.2}),
with $J_i = \{j \in J ~|~ i \in T_j \}$, for $i \in I'$.
\end{lemma}

\begdim
Suppose that
$$\C_1 = a^Iba^J + \sum_{i \in I'}a^iba^{L_i}(a-1)a^J +
\sum_{j \in J} a^I(a-1)a^{M_j}ba^j \geq 0.$$
In view of Lemma \ref{L2} (applied with $L'_i = \emptyset$),
for any $i \in I'$ there exists $k_i > 0$
such that $a^{L_{i}}(a-1)a^{J} + k_i a^{J} \geq 0$.

Assume that there exist $j \in J$ and $h \in \N$ such that
$(a^{M_{j}}(a-1)a^{I} + a^{I}, a^h) < 0$. Thus, we have
$(a^Iba^J + \sum_{j \in J} a^I(a-1)a^{M_{j}}b a^j, a^hba^j) < 0$
and so
$(\sum_{i \in I'}a^iba^{L_i}(a-1)a^J, a^hba^j) > 0$,
since $\C_1 \geq 0$.
Hence, $h \in I'$ and $(a^{L_h}(a-1)a^J, a^j) > 0$.
Consequently,
$(a^{L_h}(a-1)a^J + a^J, a^j) \geq 2$, in contradiction
with Proposition \ref{CO1}.
This proves Eq.~(\ref{3.1}).

Set, as in Eq.~(\ref{3.1}), $a^{T_j} = a^{M_{j}}(a-1)a^{I} + a^{I}$.
By Proposition \ref{Ul},
$T_j$ is a subset of $\N$ and we have
$$\C_1 = \sum_{j \in J} a^{T_j} ba^j + \sum_{i \in I'}a^iba^{L_i}(a-1)a^J
\geq 0.$$
For any $i \in I'$ such that $L_i$ is nonempty, let $\ell_i = \min L_i$. Thus,
$(\sum_{i \in I'}a^iba^{L_i}(a-1)a^J, a^i ba^{\ell_i}) < 0$
and
$$0 \leq (\C_1, a^i ba^{\ell_i}) = (\sum_{j \in J} a^{T_j} ba^j + a^iba^{L_i}(a-1)a^J,
a^i ba^{\ell_i}) < (\sum_{j \in J} a^{T_j} ba^j, a^i ba^{\ell_i}).$$
Hence, $\{i \in I' ~|~ L_i \not = \emptyset \} \subseteq \cup_{j \in J} T_j$,
i.e., Eq.~(\ref{3.0}) holds.
If we set $L_i = \emptyset$ for $i \in \cup_{j \in J} T_j \setminus I'$
and
$J_i = \{j \in J ~|~ i \in T_j \}$, we have
$$\C_1 = \sum_{j \in J} a^{T_j} ba^j +
\sum_{i \in I'}a^iba^{L_i}(a-1)a^J =
\sum_{i \in \cup_{j \in J} T_j} a^i b(a^{L_i}(a-1)a^J + a^{J_i})
\geq 0.$$
The above relation proves Eq.~(\ref{3.2}).

Conversely, let $(I,J)$ be a Krasner factorization of $\Z_{n}$
and assume that $I', L_i, M_j$  satisfy Eqs.~(\ref{3.1})--(\ref{3.2}).
If we set $L_i = \emptyset$ for $i \in \cup_{j \in J} T_j \setminus I'$
and
$J_i = \{j \in J ~|~ i \in T_j \}$, we have
\begin{eqnarray*}
\C_1 & = & a^Iba^J + \sum_{i \in I'}a^iba^{L_i}(a-1)a^J +
\sum_{j \in J} a^I(a-1)a^{M_j}ba^j \\
         & = & \sum_{j \in J} a^{T_j} ba^j + \sum_{i \in I'}a^iba^{L_i}(a-1)a^J
          =   \sum_{i \in \cup_{j \in J} T_j} a^i b(a^{L_i}(a-1)a^J + a^{J_i}) \geq 0
\end{eqnarray*}
and the proof is complete.
\enddim

\section{Construction of factorizing codes with
$S = a^J + \sum_{j \in J} a^{M_j}ba^j$} \label{constructionC}

In this section, we give a recursive characterization of positive
factorizations $(P,S)$
such that $S = a^J + \sum_{j \in J} a^{M_j}ba^j$.
Consider again the finite maximal code $\C$ defined
in Example \ref{EX1} by the relation
$\C  = P(\A-1)S + 1$, with
\begin{eqnarray*}
P & = & 1 + a^{2}ba^{\{0,1,2,3,4,5,6\}} + a^{2}ba^3ba^{\{0,1,2,3,4,5,6\}}, \\
S & = & a^{\{0,1,2,3,4\}} + a^{\{ 0,1 \}}ba^{\{0,1,2,3,4\}}.
\end{eqnarray*}
Set $P' = P_0 + P_1 = 1 + a^{2}ba^{\{0,1,2,3,4,5,6\}}$.
Equalities defining $\C_0$, $\C_1$, $\C_2$ in Example \ref{EX1},
show that $\C' = P'(\A-1)S + 1$
is a $3$-code. Moreover, $\C = \C' + zba^{L_z}(\A-1)S$,
with $zba^{L_z} = a^{2}ba^3ba^{\{0,1,2,3,4,5,6\}}$
and $zba^t = a^{2}ba^3ba^t \in C'$, for $t \in \{0,1,2,3,4\}$.
This construction can be easily generalized and
allows us to construct all factorizing
codes with $S =  a^J + \sum_{j \in J} a^{M_j}ba^j$
(Theorem \ref{teoC}).
Corollary \ref{corollario} shows that these codes may be recursively
constructed.
Proposition \ref{CO1} and Lemma \ref{Le1C} are the main tools we need in the
proof. Lemma \ref{Le2C} is a preliminary step.

\begin{lemma} \label{Le1C}
Let $(P,S)$ be a factorization for a finite
maximal code $C$,
with $S = a^J + \sum_{j \in J} a^{M_j}ba^j \in \N \langle A \rangle$,
$P = P_0 + \ldots + P_k \in \N \langle A \rangle$, $k \geq 0$.
Then $C_{k+1} \subseteq \{wba^j ~|~ w \in A^*, j \in J \}$.
\end{lemma}

\begdim
The conclusion is a direct consequence of the following
equation, where it is understood that $P_{k-1} = 0$ for $k = 0$.
$$\C_{k+1} = P_kba^J +  \sum_{j \in J}P_{k-1}ba^{M_j}ba^j
+ \sum_{j \in J}P_{k}(a-1)a^{M_j}ba^j.$$
\enddim

\begin{lemma} \label{Le2C}
Let $P, P', S$ be polynomials in $\Z \langle A \rangle$,
with $P = P_0 + \ldots + P_k + P_{k+1}$,
$P' = P_0 + \ldots + P_k$, $k \geq 0$ and
$\supp(S) \subseteq  a^* \cup a^*ba^*$.
Set $X = P(\A-1)S + 1$ and $Y = P'(\A-1)S + 1$.
Then $X_i = Y_i$ for $i \in \{0, \ldots , k \}$,
and moreover,
\begin{eqnarray*}
X_{k+1} & = & Y_{k+1} + P_{k+1}(a-1)S_0, \\
X_{k+2} & = & Y_{k+2} + P_{k+1}(a-1)S_1 + P_{k+1}bS_0, \\
X_{k+3} & = & P_{k+1}bS_1.
\end{eqnarray*}
\end{lemma}

\begdim
The conclusion is a direct consequence of Eqs.~(\ref{eqCodes0}), (\ref{eqCodesR}).
\enddim

\begin{theorem} \label{teoC}
Let $(I,J)$ be a Krasner factorization of $\Z_{n}$,
let $P, S$ be polynomials in $\N \langle A \rangle$,
with $S = a^J + \sum_{j \in J} a^{M_j}ba^j$.
Set $P' = P_0 + \ldots + P_k$ and
$P = P' + P_{k+1}$,
where $k \geq 0$ and $P_{k+1}$ is a nonnull polynomial.
Then $(P,S)$ is a positive factorization for a $(k+3)$-code
if and only if the following conditions are satisfied:
\begin{itemize}
\item[(1)]
$(P',S)$ is a positive factorization for a $(k+t)$-code $C'$, where
$t = 2$ if $P_k \not = 0$, $t = 1$ otherwise.
\item[(2)]
$P_{k+1} = \sum_{z \in Q_k} z b a^{L_{z,k+1}}$, where
$Q_k = \{z ~|~ \exists j \in J ~ zba^j \in C'_{k+1} \}$ and
$L_{z,k+1}$ is a finite subset of $\N$, for any $z \in Q_k$.
\item[(3)]
For any $z \in Q_k$, set
$J_z = \{j \in J ~|~ zba^j \in C'_{k+1} \}$. We have
\begin{eqnarray} \label{Eq12dir}
 \forall z \in Q_{k} \quad a^{L_{z,k+1}}(a-1)a^J + a^J \geq
a^{L_{z,k+1}}(a-1)a^J + a^{J_z} \geq 0, & & \\
\label{Eq13dir}
\forall j \in J, z \in Q_{k} \setminus \supp(P_k) \quad a^{L_{z,k+1}}(a-1)a^{M_j} +
a^{L_{z,k+1}} \geq 0. & &
\end{eqnarray}
\end{itemize}
\end{theorem}

\begdim
Let $(I,J)$ be a Krasner factorization of $\Z_{n}$.
Let $P, P', S$ be polynomials in $\N \langle A \rangle$,
where $P = P_0 + \ldots + P_k + P_{k+1}$,
$P' = P_0 + \ldots + P_k$, $k \geq 0$, $P_{k+1}$ is a
nonnull polynomial and
$S = a^J + \sum_{j \in J} a^{M_j}ba^j$.

Suppose that $(P',S)$ is a positive factorization for a $(k+t)$-code $C'$,
with $t = 2$ if $P_k \not = 0$, $t = 1$ otherwise,
$P_{k+1}$ is as in item (2) and Eqs.~(\ref{Eq12dir}), (\ref{Eq13dir}) are satisfied.
By Lemma \ref{Le1C}, $C'_{k+1} \subseteq \{wba^j ~|~ w \in A^*, j \in J \}$.
Thus, $\C'_{k+1} = \sum_{z \in Q_k} z b a^{J_z}$.
Let us prove that $(P,S)$ is a positive factorization for a $(k+3)$-code $C$.
By Lemma \ref{Le2C}, since
$P_{k+1}bS_1 = \sum_{j \in J}P_{k+1} b a^{M_j}ba^j \geq 0$,
it suffices to prove that the following relations hold
\begin{eqnarray} \label{eqGen1}
\C_{k+1} & = & \C'_{k+1} + P_{k+1}(a-1)a^J \geq 0, \\
\label{eqGen2}
\C_{k+2} & = & \C'_{k+2} +  \sum_{j \in J} P_{k+1}(a-1)a^{M_j}ba^j + P_{k+1}ba^J \geq 0.
\end{eqnarray}
By Eq. (\ref{Eq12dir}) we have
$$\C_{k+1} = \C'_{k+1}  + P_{k+1}(a-1)a^J = \C'_{k+1} +  \sum_{z \in Q_k} z b a^{L_{z,k+1}}(a-1)a^J
= \sum_{z \in Q_{k}} z b(a^{J_z} + a^{L_{z,k+1}}(a-1)a^J) \geq 0,$$
hence Eq.~(\ref{eqGen1}) is satisfied. Next, we have
\begin{eqnarray*}
\C_{k+2} &= & \C'_{k+2} + \sum_{j \in J} P_{k+1}(a-1)a^{M_j}ba^j + P_{k+1}ba^J \\
&=& \sum_{j \in J} P_{k}b a^{M_j}ba^j  +
\sum_{j \in J, \;z \in Q_k}z b (a^{L_{z,k+1}}(a-1)a^{M_j} + a^{L_{z,k+1}})ba^j  \geq \\
& & \sum_{j \in J, \; z \in Q_k \cap \supp(P_k)}zb (a^{M_j} + a^{L_{z,k+1}}(a-1)a^{M_j} + a^{L_{z,k+1}})ba^j \\
& + &  \sum_{j \in J, \; z \in Q_k \setminus \supp(P_k)}zb (a^{L_{z,k+1}}(a-1)a^{M_j} + a^{L_{z,k+1}})ba^j
\end{eqnarray*}
In turn, by Eq.~(\ref{Eq13dir}) we have
$$\sum_{j \in J, \; z \in Q_k \setminus \supp(P_k)}zb (a^{L_{z,k+1}}(a-1)a^{M_j} + a^{L_{z,k+1}})ba^j
\geq 0,$$
then, by Lemma \ref{LG1}, $a^{M_{j}}(a-1)a^{I} + a^{I} \geq 0$
for any $j \in J$, and by Proposition \ref{holes} we have
$$\sum_{j \in J, \; z \in Q_k \cap \supp(P_k)}zb (a^{M_j} + a^{L_{z,k+1}}(a-1)a^{M_j} + a^{L_{z,k+1}})ba^j
\geq 0.$$
Therefore, Eq.~(\ref{eqGen2}) is also satisfied and $(P,S)$ is a positive factorization for a $(k+3)$-code $C$.

Conversely, assume that $(P,S)$ is a positive factorization for a $(k+3)$-code $C$.
Set $P_{k+1} = \sum_{z \in Q_k} z b a^{L_{z,k+1}}$,
where $Q_k$
is a finite set of words and $L_{z,k+1}$ is a finite nonempty subset of $\N$, for any $z \in Q_k$.

We first prove that the polynomial
$Y_{k+1} = P_kba^J + \sum_{j \in J} P_{k-1} b a^{M_j}ba^j +
\sum_{j \in J} P_k (a-1) a^{M_j}ba^j$ is in $\N \langle A \rangle \setminus \{0 \}$.
Notice that if $wba^j \in \supp(Y_{k+1})$, then $j \in J$.
By hypothesis, $\C_{k+1} =  Y_{k+1} + \sum_{z \in Q_k} z b a^{L_{z,k+1}}(a-1)a^J  \geq 0$.
Moreover, we have $(a^{L_{z,k+1}}(a-1)a^J, a^q) < 0$, for $q = \min L_{z,k+1}$.
For any $q \in \N$ such that
$(a^{L_{z,k+1}}(a-1)a^J, a^q) < 0$ we get
$(P_{k+1}(a-1)a^J, zba^q) < 0$. Thus, since $\C_{k+1} \geq 0$,
we also have $(Y_{k+1} , zba^q) > 0$, that is $q \in J$.
As a consequence, $a^{L_{z,k+1}}(a-1)a^J + ka^J \geq 0$,
where
$$k = \max \{h ~|~ (a^{L_{z,k+1}}(a-1)a^J, a^q) = - h, ~ q \in J \} > 0.$$
If there existed $w \in A^*$ and $j \in \N$ such that $(Y_{k+1} , wba^j) < 0$,
since $\C_{k+1} \geq 0$, we should have
$( \sum_{z \in Q_k} z b a^{L_{z,k+1}}(a-1)a^J, wba^j) > 0$.
Thus, there should be $z \in Q_k$ such that $(a^{L_{z,k+1}}(a-1)a^J, a^j) > 0$.
Moreover, the word $wba^j$ is in $\supp(Y_{k+1})$, hence, as noticed
before, $j \in J$ and so $(a^{L_{z,k+1}}(a-1)a^J + a^J, a^j) \geq 2$.
This relation is in contradiction
with Proposition \ref{CO1}.
In conclusion, $Y_{k+1} \in \N \langle A \rangle \setminus \{0 \}$. Of course,
$Y_{k+2} = \sum_{j \in J} P_kba^{M_j}ba^j \geq 0$. Finally,
$\C_i = Y_i \geq 0$, for $i \in \{0, \ldots , k \}$ (Lemma \ref{Le2C}).
Hence, $(P',S)$ is a positive factorization
for a $(k+t)$-code $C'$, with
$t = 2$ if $P_k \not = 0$, $t = 1$ otherwise.
In turn, this implies $Y_{k+1} = \C'_{k+1}$
and, by Lemma \ref{Le1C}, we may set $\C'_{k+1} = \sum_{z \in Q'_k} zba^{J_z}$,
where $J_z$ is a nonempty subset of $J$.

We have already observed that for any $z \in Q_k$,
we have $(\C'_{k+1}, zba^{\ell}) > 0$ for $\ell = \min L_{z,k+1}$.
Therefore, $Q_k \subseteq Q'_k$
and we may assume $Q_k = Q'_k$
if we define $L_{z,k+1} = \emptyset$ for any word $z \in Q'_k \setminus Q_k$. Therefore,
condition (2) holds.
Then by
$$\C_{k+1} =  \sum_{z \in Q_k} zba^{J_z} + \sum_{z \in Q_k} z b a^{L_{z,k+1}}(a-1)a^J  \geq 0$$
Eq.~(\ref{Eq12dir}) easily follows.
Finally, by
$$\C_{k+2}  = \sum_{j \in J}P_{k}ba^{M_j}ba^j  +
 \sum_{z \in Q_k, j \in J} z b a^{L_{z,k+1}}(a-1)a^{M_j}ba^j + \sum_{z \in Q_k}z b a^{L_{z,k+1}}ba^J \geq 0$$
Eq.~(\ref{Eq13dir}) easily follows.
\enddim

\begin{corollary} \label{corollario}
Let $(I,J)$ be a Krasner factorization of $\Z_{n}$,
let $P, S$ be polynomials in $\N \langle A \rangle$,
with $S = a^J + \sum_{j \in J} a^{M_j}ba^j$.
Set $P = P_0 + \ldots + P_{k+1}$,
with $k \geq 0$.
If $(P,S)$ is a positive factorization for a $(k+3)$-code $C$,
then $(P_0 + \ldots + P_r, S)$ is a positive factorization
for a $(r+2)$-code for any $r \in \{0, \ldots , k \}$ such that $P_r \not = 0$.
\end{corollary}

\begdim
The conclusion may be easily obtained by using induction and Theorem
\ref{teoC}.
\enddim

\section{Construction of $4$-codes} \label{construction4}

In this section we focus on positive factorizations $(P,S)$
for $4$-codes $C$ satisfying item (3) in Lemma \ref{M0}
and such that $I' = I$, i.e.,
\begin{eqnarray} \label{pair}
P & = & a^I + \sum_{i \in I}a^iba^{L_i} +
\sum_{i \in I, \ell \in L_i}a^iba^{\ell}ba^{L_{i,\ell}}, \quad
S =  a^J + \sum_{j \in J'} a^{M_{j}}b a^j,
\end{eqnarray}
where
$(I,J)$ is a Krasner factorization of $\Z_{n}$ and
$J', L_i, M_j, L_{i,\ell}$ are finite subsets of $\N$,
for any $i,j, \ell$.
We give a characterization of these pairs.
Their first property is stated by Lemma \ref{4CL1}.
This lemma, which is true also when $I' \not = I$, shows that the following relations
hold:
\begin{eqnarray} \label{83}
\forall  i \in I, \ell \in L_i,  j \in J' \cap J
 & & a^{L_{i,\ell}}(a-1)a^{M_j} + a^{L_{i,\ell}} + a^{M_j} \geq 0,
 \end{eqnarray}
 \begin{eqnarray} \label{84}
 \forall  i \in I, \ell \in L_i, j \in J' \setminus J
 & &  a^{L_{i,\ell}}(a-1)a^{M_j} + a^{M_j} \geq 0.
 \end{eqnarray}

\begin{lemma} \label{4CL1}
Let $I, J', J, L_i, M_j, L_{i,\ell}$ be finite subsets of $\N$,
for any $i,j, \ell$. We have
\begin{eqnarray*}
X_3 & = & \sum_{i \in I, \ell \in L_i}a^iba^{\ell}ba^{L_{i,\ell}}ba^J +
\sum_{i \in I, j \in J'}a^iba^{L_i}ba^{M_j}ba^j
\end{eqnarray*}
\begin{eqnarray*}
& + &
\sum_{i \in I, \ell \in L_i, j \in J'}a^iba^{\ell}ba^{L_{i,\ell}}(a-1)a^{M_j}ba^j \geq 0
\end{eqnarray*}
if and only if Eqs.~(\ref{83}), (\ref{84}) hold.
\end{lemma}

\begdim
Let us write the polynomial $X_3$ in a different way:
$$X_3 = \sum_{i \in I, \ell \in L_i, j \in J \setminus J'}a^iba^{\ell}ba^{L_{i,\ell}}ba^j
 +   \sum_{i \in I, \ell \in L_i, j \in J' \cap J}a^iba^{\ell}b
 (a^{L_{i,\ell}}+a^{M_j} + a^{L_{i,\ell}}(a-1)a^{M_j}) ba^j $$
$$ + \sum_{i \in I, \ell \in L_i, j \in J' \setminus J}a^iba^{\ell}b
 (a^{L_{i,\ell}}(a-1)a^{M_j} + a^{M_j}) ba^j .$$
Then $X_3$ is a polynomial in $\N \langle A \rangle$ if and only if
the second and the third sum on the right side
of the above equation are also polynomials in
$\N \langle A \rangle$. Hence Eqs.~(\ref{83}), (\ref{84}) easily follow.
 \enddim

Let $(P,S)$ be a factorization for $C$ satisfying Eq.~(\ref{pair}). Then
Lemma \ref{LG1} applies to the factorization $(S^{\sim}, P^{\sim})$
of $C^{\sim}$ and
Eqs.~(\ref{3.1})--(\ref{3.2}) become:
\begin{eqnarray} \label{Eq81}
\forall i \in I & & a^{R_i} = a^{L_{i}}(a-1)a^{J} + a^{J} \geq 0,
\end{eqnarray}
\begin{eqnarray} \label{Eq80}
& & \{j \in J' ~|~ M_j \not = \emptyset \} \subseteq \cup_{i \in I} R_i,
\end{eqnarray}
\begin{eqnarray} \label{Eq82}
\forall j \in J'& & a^{T_{j}} = a^{M_{j}}(a-1)a^{I} + a^{I} \geq
a^{M_{j}}(a-1)a^{I} + a^{I_j} \geq 0,
\end{eqnarray}
where $I_j = \{i \in I ~|~ j \in R_i \}$, for $j \in J'$.
Proposition \ref{direct4I} shows that two further relations
are required in order to characterize this family of positive
factorizations for a finite maximal code.

\begin{proposition} \label{direct4I}
Let $P = a^I + \sum_{i \in I} a^iba^{L_i} +
\sum_{i \in I, \ell \in L_i}a^iba^{\ell}ba^{L_{i,\ell}}$,
$S = a^J + \sum_{j \in J'} a^{M_{j}}b a^j$, where
$(I,J)$ is a Krasner factorization of $\Z_{n}$,
$J', L_i, M_j, L_{i,\ell}$ are finite subsets of $\N$,
for any $i,j, \ell$. Then $(P,S)$ is a positive
factorization for a finite maximal code $C$ if and only if
Eqs.~(\ref{83})--(\ref{Eq82})
hold and, moreover,
\begin{eqnarray} \label{85bis}
\forall  i \in I, \ell \in L_{i} & &
a^{R_{i,\ell}} = a^{L_{i,\ell}}(a-1)a^J + a^J \geq 0,
\end{eqnarray}
\begin{eqnarray} \label{86bis}
\forall i \in I, j  \in J', \ell \in L_i& &
(a^{L_i}(a-1)a^{M_j} + a^{M_j}, a^{\ell}) < 0 \quad \Rightarrow \quad j \in  R_{i,\ell}.
\end{eqnarray}
\end{proposition}

\begdim
Assume that $P,S$ are as in the statement and
Eqs.~(\ref{83})--(\ref{86bis})
hold.
Since $L_i$, $M_j$ satisfy Eqs.~(\ref{Eq81})--(\ref{Eq82}),
for any $i \in I$, $j \in J'$, the polynomial
$a^{L_i}(a-1)a^{M_j} + a^{L_i} + a^{M_j} $
has coefficients $0,1$
(Proposition \ref{holes}). Therefore,
if $(a^{L_i}(a-1)a^{M_j} + a^{M_j}, a^{\ell}) < 0$ then
$\ell \in L_i$ and $(a^{L_i}(a-1)a^{M_j} + a^{M_j} + a^{\ell}, a^{\ell}) \geq 0$.

Let us show that $(P,S)$ is a positive factorization for a code $C$,
i.e., $P(A-1)S + 1 = \C \geq 0$.
We have to prove that $\C_h \geq 0$ for $h \in \{0, 1, 2, 3, 4 \}$.
Of course, $\C_0 = a^I(a-1)a^J + 1 = a^n \geq 0$
and $\C_4 = \sum_{i \in I, \ell \in L_i, j \in J'}a^iba^{\ell}ba^{L_{i,\ell}} b a^{M_{j}}b a^j \geq 0$.
Furthermore, since Eqs.~(\ref{Eq81})--(\ref{Eq82}) hold,
we have $\C_1 \geq 0$ (Lemma \ref{LG1} applied to $\C^{\sim}_1$) and since
Eqs.~(\ref{83}),  (\ref{84}) hold, we have $\C_3 \geq 0$
(Lemma  \ref{4CL1}).
Finally, in view of Eqs.~(\ref{85bis}),
(\ref{86bis}) we have
\begin{eqnarray*}
\C_2 & = & \sum_{i \in I}a^iba^{L_i}ba^J +
\sum_{j \in J'} a^I b a^{M_j}ba^j  +
\sum_{i \in I, j \in J'}a^iba^{L_i}(a-1) a^{M_j}ba^j \\
& + &
\sum_{i \in I, \ell \in L_i}a^iba^{\ell}ba^{L_{i,\ell}}(a-1)a^J \\
& = & \sum_{i \in I, \ell \in L_i}a^iba^{\ell}ba^{R_{i,\ell}} +
\sum_{i \in I, j \in J'}a^ib(a^{L_i}(a-1) a^{M_j} + a^{M_j})ba^j \geq 0.
\end{eqnarray*}

Conversely, let $P = a^I + \sum_{i \in I} a^iba^{L_i} +
\sum_{i \in I, \ell \in L_i}a^iba^{\ell}ba^{L_{i,\ell}}$,
$S = a^J + \sum_{j \in J'} a^{M_{j}}b a^j$, where
$(I,J)$ is a Krasner factorization of $\Z_{n}$,
$J', L_i, M_j, L_{i,\ell}$ are finite subsets of $\N$,
for any $i,j, \ell$.
Assume $P(A-1)S + 1 = \C \geq 0$, thus
$\C_h \geq 0$ for $h \in \{0, 1, 2, 3, 4 \}$.
Since $\C_1 \geq 0$, by Lemma  \ref{LG1} applied to $\C^{\sim}_1$,
Eqs.~(\ref{Eq81})--(\ref{Eq82}) hold.
Hence, by Proposition \ref{holes},  for any $i \in I$, $j \in J'$,
$a^{L_i}(a-1)a^{M_j} + a^{L_i} + a^{M_j}$
is a polynomial with coefficients $0,1$.
In addition, since  $\C_3 \geq 0$,
Eqs.~(\ref{83}),  (\ref{84}) hold (Lemma  \ref{4CL1}).
Furthermore, we have
\begin{eqnarray*}
\C_2 & = & \sum_{i \in I}a^iba^{L_i}ba^J +
\sum_{j \in J'} a^I b a^{M_j}ba^j  +
\sum_{i \in I, j \in J'}a^iba^{L_i}(a-1) a^{M_j}ba^j
\end{eqnarray*}
\begin{eqnarray*}
& + &
\sum_{i \in I, \ell \in L_i}a^iba^{\ell}ba^{L_{i,\ell}}(a-1)a^J \geq 0.
\end{eqnarray*}
Assume that there are $i \in I$, $\ell \in L_{i}$, $t \in \N$ such that
$(a^{L_{i,\ell}}(a-1)a^J + a^J , a^t) < 0$. Thus,
$$(\sum_{i \in I}a^iba^{L_i}ba^J + \sum_{i \in I, \ell \in L_i}a^iba^{\ell}ba^{L_{i,\ell}}(a-1)a^J,
a^i b a^{\ell}ba^t) < 0.$$
Since $\C_2 \geq 0$, we have
$$(\sum_{j \in J'} a^I b a^{M_j}ba^j  +
\sum_{i \in I, j \in J'}a^iba^{L_i}(a-1) a^{M_j}ba^j, a^i b a^{\ell}ba^t) > 0$$
which yields $( a^{M_t} + a^{L_i}(a-1) a^{M_t}, a^{\ell}) > 0$, i.e.,
$( a^{M_t} + a^{L_i}(a-1) a^{M_t} + a^{L_i}, a^{\ell}) > 1$. The last relation
is impossible since $a^{L_i}(a-1)a^{M_t} + a^{L_i} + a^{M_t}$
is a polynomial with coefficients $0,1$.
Therefore, Eq.~(\ref{85bis}) holds.
Finally, we have
\begin{eqnarray*}
\C_2 & = & \sum_{i \in I, \ell \in L_i}a^iba^{\ell}ba^{R_{i,\ell}} +
\sum_{i \in I, j \in J'}a^ib(a^{L_i}(a-1) a^{M_j} + a^{M_j})ba^j \geq 0.
\end{eqnarray*}
By the above equation, for all $i \in I$, $j  \in J'$, $\ell \in L_i$ such that
$(a^{L_i}(a-1)a^{M_j} + a^{M_j}, a^{\ell}) < 0$, we have
$j \in  R_{i,\ell}$, i.e., Eq.~(\ref{86bis}) holds.
\enddim

Looking at Proposition \ref{direct4I},  we see that two cases
may occur: either for any
$i \in I$ and $j  \in J'$ we have
$a^{L_i}(a-1)a^{M_j} + a^{M_j}  \geq 0$ or not.
Examples \ref{SIC} and
\ref{NOC1} from \cite{DF06} illustrate the first and the second
case respectively and point out relations between $4$-codes, $3$-codes
and Haj\'{o}s factorizations.

\begin{example} \label{SIC}
{\rm Let us
consider
the polynomials:
\begin{eqnarray*}
P &=& a^{\{0,2,4,12,14,16\}}+
a^{\{0,2,4,12,14,16\}}ba^{\{ 1,3,5,7,9,11,13,15,17,19 \}} + \\
 & & a^{\{0,2,4,12,14,16\}}ba^{\{ 1,3,5,7,9,11,13,15,17,19 \}}
ba^{\{ 1,3,5,7,9,11,13,15,17,19 \}},\\
S &=& a^{\{0,1,6,7\}} + a^{\{2,3\}}ba^{21}.
\end{eqnarray*}

An easy computation shows that $(P,S)$ is a positive
factorization for a $4$-code
$\C = P(A -1)S + 1$.
The pair $(I,J) = (\{0,2,4,12,14,16\}, \{0,1,6,7\})$ is a Krasner
factorization of $\Z_{24}$. Moreover,
$L_i = L_{i,\ell} = L = \{ 1,3,5,7,9,11,13,15,17,19 \}$, $M_j = M = \{2,3\}$,
$J' = \{21\}$.
There is a strong Haj\'{o}s factorization $(T,R)$
of $\Z_{24}$ associated with $(P,S)$, namely
$(T, R)=(\{0,4,8,12,16,20\}, \{0,27,6,21 \})$,
with the corresponding pair
$(M,L') = (\{2,3\}, \{ 1,3,5,7,9,11,$ $13,15,17,19 \})$.
It is easy to see that
the pair $(P',S)$, with
$P' = P_0 + P_1$,
defines a $3$-code.}
\end{example}

\begin{example} \label{NOC1}
{\rm Let us
consider
the polynomials:
\begin{eqnarray*}
P &=& a^{\{0,2,4,12,14,16\}}+
a^{\{0,2,4,12,14,16\}}ba^{\{ 1,9,11,13 \}} + \\
 & & a^{\{0,2,4,12,14,16\}}ba^{\{ 1,9,11,13 \}}
ba^{\{ 1,3,5,7,9,11,13,15,17,19 \}},\\
S &=& a^{\{0,1,6,7\}} + a^{\{2,3\}}ba^{21}.
\end{eqnarray*}
An easy computation shows that $(P,S)$ is a positive
factorization for a $4$-code
$\C = P(A -1)S + 1$.
We have
$L_i = L = \{ 1,9,11,13 \}$, $M_j = M = \{2,3\}$,
$J' = \{21\}$, $L_{i,\ell} = L' = \{ 1,3,5,7,9,11,13,15,17,19 \}$.
There are two Haj\'{o}s factorizations
of $\Z_{24}$ associated with $(P,S)$:
the strong Haj\'{o}s factorization
$(T, R')=(\{0,4,8,12,16,20\}, \{0,27,6,21 \})$
with the corresponding pair
$(M,L') = (\{2,3\}, \{ 1,3,5,7,9,11,13,15,17,19 \})$
and the non-strong Haj\'{o}s factorization
$(T, R)=(\{0,4,8,12,16,20\}, \{0,3,6,21 \})$
with the corresponding
pair $(M,L) = (\{2,3\}, \{1,9,11,13\})$.
Notice that $R \cap R' = \{21 \}$.
The pair $(P', S)$, with
$P' = P_0 + P_1$,
does not define a $3$-code.}
\end{example}

\section{Conclusions} \label{finale}

In this paper we proved that if $(P,S)$
is a factorization for a finite maximal code $C$,
with
$C \cap a^* = \{ a^p \}$
for a prime number $p$,
$S = S_0 + S_1$ and if $a^j \in \supp(S)$
for any $a^iba^j \in \supp(S)$,
then $(P,S)$ is positive.
We also proved that $(p,4)$-codes satisfy
the factorization conjecture and each factorization $(P,S)$
for a $(p,4)$-code is positive.

A natural question is to characterize
those positively factorizing codes having only positive
factorizations. One may conjecture that this is the case
for finite maximal codes
containing a power of $a$ with a prime exponent.
This is a first research direction.

A related problem is to find conditions under which
a factorization is positive. In this framework,
Hansel and Krob asked the following question, reported in \cite{KH}:
let $P, Q \in \Z[a]$ be such that $PQ = 1 + a + \ldots + a^n$,
with $n \in \N$. If $P$ has coefficients $0,1$ then does
$Q$ also have coefficients $0,1$?

As a second direction of research for both the above mentioned
problems, one can investigate finite maximal codes
containing a power of $a$ with a prime exponent
and having a factorization $(P,S)$, where
$S = S_0 + S_1 + \ldots + S_k$ is a polynomial such that
if $a^ibw \in \supp(S_j)$, then $w \in \supp(S_{j-1})$, for
$j \in \{1, \ldots , k \}$. One may ask whether it is still true that
$P,S$ are necessarily in $\N \langle A \rangle$.
We have already proved that this statement holds if
$S$ is a polynomial in $\N \langle A \rangle$ such that
$a^j \in \supp(S)$,
for any $wba^j$ in $\supp(S)$
(Theorem \ref{mainteoremGen}).

Concerning the structure of positively factorizing codes,
our construction of $(p,4)$-codes is not complete.
More generally, a method for constructing all positively factorizing codes
is still lacking (see \cite{DF06,DF07a,DF07b}
for conjectures and related problems).
In this regard, it could be interesting to look for
a generalization of the constructions given in Sections
\ref{constructionC}, \ref{construction4}.

Another related question is to find conditions under which
a set of words $C_1$ satisfying
$\C_1 = a^I(a-1)S_1 +  P_1(a-1)a^J + a^Iba^J$, where
$(I,J)$ is a Krasner pair and $P_1, S_1$ are polynomials
with coefficients $0,1$, could be embedded in a factorizing code.
Some sufficient conditions have been stated in
\cite{DF96,DF05,DF07a,DF07b}.

Finally, one can investigate whether all the results concerning
$m$-codes can be generalized to alphabets having
cardinality greater than two.
%--------------------------------

\paragraph{Acknowledgments}
The author thanks the anonymous referee for his/her helpful
suggestions and constructive criticism.

The final publication is available at http://link.springer.com
(Acta Informatica,  DOI: 10.1007/s00236-013-0187-1) 	

%--------------------------

\end{document}